\documentclass[article,superscriptaddress,10pt,showkeys,showpacs,twocolumn]{revtex4}
\usepackage{amsfonts}
\usepackage{amsmath}
\usepackage{amssymb}
\usepackage{graphicx}
\usepackage{longtable}

\begin{document}

\title{Photon blockade in a double-cavity optomechanical system with nonreciprocal coupling}

\author{Dong-Yang Wang}
\affiliation{School of Physics, Harbin Institute of Technology, Harbin, Heilongjiang 150001, China}
\author{Cheng-Hua Bai}
\affiliation{School of Physics, Harbin Institute of Technology, Harbin, Heilongjiang 150001, China}
\author{Shutian Liu}
\email{stliu@hit.edu.cn}
\affiliation{School of Physics, Harbin Institute of Technology, Harbin, Heilongjiang 150001, China}
\author{Shou Zhang}
\email{szhang@ybu.edu.cn}
\affiliation{Department of Physics, College of Science, Yanbian University, Yanji, Jilin 133002, China}
\author{Hong-Fu Wang}
\email{hfwang@ybu.edu.cn}
\affiliation{Department of Physics, College of Science, Yanbian University, Yanji, Jilin 133002, China}


\date{\today}

\begin{abstract}
Photon blockade is an effective way to generate single photon, which is of great significance in quantum state preparation and quantum information processing. Here we investigate the statistical properties of photons in a double-cavity optomechanical system with nonreciprocal coupling, and explore the photon blockade in the weak and strong coupling regions respectively. To achieve the strong photon blockade, we give the optimal parameter relations under different blockade mechanisms. Moreover, we find that the photon blockades under their respective mechanisms exhibit completely different behaviors with the change of nonreciprocal coupling, and the perfect photon blockade can be achieved without an excessively large optomechanical coupling, i.e., the optomechanical coupling is much smaller than the mechanical frequency, which breaks the traditional cognition. Our proposal provides a feasible and flexible platform for the realization of single-photon source.
\end{abstract}
\maketitle

\section{\label{sec.1}Introduction}
Photon blockade (PB) is a nonclassical anti-bunching effect and satisfies sub-Poissonian light statistics, which is the crucial role to generate and manipulate the single-photon source in most quantum information and computation sciences, such as quantum simulation~\cite{PhysRevLett.105.137401}, quantum key distribution~\cite{RevModPhys.81.1301}, quantum repeater~\cite{PhysRevA.72.052330}, quantum metrology~\cite{PhysRevLett.114.170802}, etc. So far, there are two well-known ideas to achieve the strong PB effect, i.e., the conventional photon blockade (CPB)~\cite{PhysRevLett.79.1467,PhysRevA.49.R20,PhysRevA.46.R6801} and the unconventional photon blockade (UPB)~\cite{PhysRevLett.104.183601,PhysRevA.83.021802,NJP.15.025014} mechanisms, which respectively rely on the anharmonic eigenenergy spectrum and the destructive quantum interference between two different quantum transition paths from the ground state to a two-excited state, and both ideas have been realized experimentally in various systems~\cite{Nature.436.87,NatPhys.4.859,PhysRevLett.107.053602,PhysRevLett.121.043601,PhysRevLett.121.043602}.

Specifically, the anharmonic eigenenergy spectrum of CPB usually comes from kinds of nonlinearities, e.g., Kerr-type nonlinear dielectrics~\cite{PhysRevB.85.033303,PhysRevB.87.235319} and nonlinear coupling~\cite{PhysRevA.99.043837,JPA.51.095302}. When the external laser driving is resonant to the single-photon transition and the energy gap of the two-photon transition is larger than the cavity decay, the CPB occurs. Furthermore, the anharmonic eigenenergy spectrum can also be obtained via the dispersive coupling when the system includes a two-level qubit~\cite{PhysRevA.82.032101,PhysRevA.87.023809,PhysRevA.99.063828,PhysRevX.10.021022}. On the other hand, the UPB mechanism usually requires an extra component, such as atom~\cite{PhysRevLett.108.183601,PhysRevLett.109.193602,PhysRevLett.118.133604,PhysRevLett.122.243602,PhysRevApplied.12.044065,PhysRevA.98.023856,PhysRevA.100.033814}, magnon~\cite{PhysRevA.100.043831,PhysRevB.100.134421}, optical parametric amplifier~\cite{PhysRevA.96.053827}, qubit~\cite{OE.25.6767}, nonlinear reservoir~\cite{PhysRevA.100.053857}, auxiliary cavity~\cite{PhysRevA.95.043838,PhysRevA.92.023838,PhysRevA.91.063808,OE.23.32835}, etc, which is used to construct another transition path from the ground state to the two-excited state. The UPB is obtained when the destructive quantum interference occurs between the direct excitation path and the structured path. In the past decade, the optomechanical system has become a hot topic to study PB due to its inherent nonlinearity~\cite{PhysRevLett.107.063601,PhysRevLett.107.063602,SC.62.970311,OL.43.2050}. However, the implementation of strong PB has to require a very large optomechanical coupling strength according the CPB mechanism, which is the major obstacle to experimental realization. In addition, those researches on the UPB in optomechanical system have been proposed via inducing the auxiliary cavity mode~\cite{PhysRevLett.109.013603,PhysRevA.87.013839,PhysRevA.98.013826,JPB.46.035502,arXiv:1302.5937} or parametric amplifier~\cite{OL.45.2604}. The coupling between the auxiliary cavity and optomechanical system is usually larger than the cavity decay to achieve the complete destructive quantum interference. So how to achieve a strong PB effect in the weak coupling region will be beneficial to study in experiments. Reference~\cite{PhysRevA.96.053810} has reported an UPB effect in the weak coupling region via optimizing the two-driving relation, in which both components are driven by the tuned pump fields.

Currently, the non-Hermitian system has attracted much attention of numerous researchers due to its own properties~\cite{RepProgPhys.70.947,Science.363.eaar7709}. As the name implies, the non-Hermitian system is in violation of the Hermitian property, which can be divided into two different patterns, i.e., gain system~\cite{PhysRevLett.80.5243,IEEEJQE.46.1626,Nat.Phys.10.394,Science.346.328,PhysRevA.92.053837,PhysRevA.100.053820} and nonreciprocal coupling system~\cite{PhysRevLett.121.086803,PhysRevB.22.2099,PhysRevB.97.045106,PhysRevA.97.052115}. As an extreme case of nonreciprocal coupling, the unidirectional dissipative coupling is studied in a coupled nonlinear cavity system~\cite{PhysRevA.94.013815}. Moreover, we have also studied and distinguished the different blockade mechanisms in a $\mathcal{PT}$-symmetric optomechanical system with gain~\cite{PhysRevA.99.043818}. Naturally, we are expected to further take account of the effect of the nonreciprocal coupling on PB, and achieve the PB in non-Hermitian system. It is worth emphasizing that, what we investigate in this paper is essentially different from that of the nonreciprocal PB~\cite{PhysRevLett.121.153601,PhysRevA.100.053832,PhysRevA.101.013826,PRJ.7.630,PhysRevApplied.13.044070}, which represents whether or not the blockade occurs when the system is driven by laser from different directions.

Here, we are dedicated to investigating the PB effect in a double-cavity optomechanical system, where the coupling between the two cavities is different in direction. As reported in various theoretical and experimental proposals~\cite{PhysRevLett.75.4710,PhysRevLett.77.570,PhysRevB.92.094204,PhysRevLett.107.173902,SR.5.13376,PhysRevResearch.2.013280}, the system under consideration consists of two coupled whispering-galley-mode cavities, where one is an optomechanical cavity and the other is an ordinary cavity driven by a weak classical laser field. In our proposal, the PB effect is measured by utilizing the second-order correlation function, which is calculated by analytically solving the Schr\"odinger equation or numerically simulating the quantum master equation, respectively. Moreover, we distinguish the different blockade mechanisms (CPB or UPB) both in the weak and strong coupling regions, and discuss in detail the effects of parameter fluctuation on CPB and UPB under the steady-state assumption. We find that the CPB and UPB exhibit completely different behaviors with the change of the nonreciprocal coupling. Meanwhile, to ensure the effective implementation of strong PB, we respectively give the optimal parameter relations for the different blockade mechanisms. Furthermore, by simulating the dynamic evolution of the initial system numerically, all the analysis and discussion under the steady-state assumption are further verified. Different from the previous proposals~\cite{PhysRevLett.107.063601,PhysRevLett.107.063602,PhysRevA.88.023853,PhysRevA.92.033806,PhysRevA.96.013861,PhysRevA.99.013804,OL.43.1163,OE.27.27649,PhysRevLett.109.013603,PhysRevA.87.013839,PhysRevA.98.013826,JPB.46.035502,arXiv:1302.5937}, the perfect PB not only does not need the excessively large optomechanical coupling, but  also can be obtained with a weak nonreciprocal coupling, i.e., the optomechanical coupling is much smaller than the mechanical frequency and the nonreciprocal coupling is smaller than the cavity decay. And only one cavity is driven by a normal laser field, which is different from the two-driving tuned scheme~\cite{PhysRevA.96.053810}. Therefore, our proposal would be more feasible for the realization of strong PB, and ensures that a high-quality and efficient single-photon source can be engineered.

The paper is organized as follows: In Sec.~\ref{sec.2}, we illustrate the double-cavity optomechanical system with nonreciprocal coupling and give its Hamiltonian. In Sec.~\ref{sec.3}, we analytically and numerically solve the second-order correlation function, analyze the effects of the nonreciprocal coupling on the PB, and discuss the different blockade mechanisms from weak region to strong coupling region. Finally, a conclusion is given in Sec.~\ref{sec.4}.

\section{\label{sec.2}System and Hamiltonian}
\begin{figure}
	\includegraphics[width=0.8\linewidth]{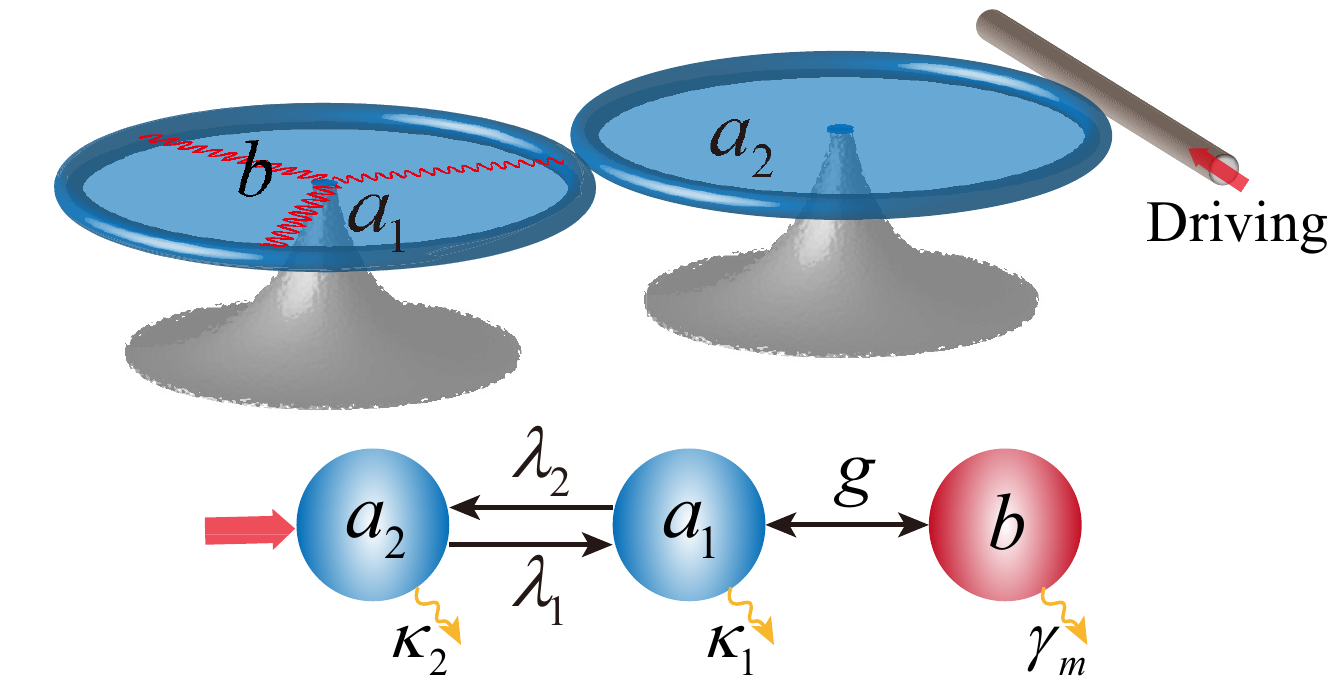}
	\caption{\label{fig:1}Schematic diagram of the double-cavity optomechanical system with nonreciprocal coupling.}
\end{figure}

As depicted in Fig.~\ref{fig:1}, we consider a double-cavity optomechanical system~\cite{PhysRevLett.113.053604,PhysRevA.80.033821,PhysRevA.90.053841,PhysRevA.91.033818,PhysRevA.92.023856,SR.6.38559} consisting of one mechanical mode ($b$) and two optical modes ($a_{1}$ and $a_{2}$), where the coupling between two cavities is nonreciprocal ($\lambda_{1}\neq\lambda_{2}$) and the mechanical mode is coupled to the optical mode $a_{1}$ via optomechanical interaction. The nonreciprocal coupling in our proposal can be achieved by utilizing an imaginary gauge field~\cite{PhysRevLett.77.570,PhysRevB.92.094204}, impurity~\cite{PhysRevLett.107.173902}, or auxiliary nonreciprocal transition device~\cite{SR.5.13376,PhysRevResearch.2.013280}. Such as, the nonreciprocal coupling will be scaled by $\exp(\pm h)$ in the opposite transmission directions when the system is placed in an imaginary gauge field, where $h$ describes the effect of the imaginary vector potential. On the other hand, the presence of impurity or auxiliary nonreciprocal transition device can also affect the photon hopping between cavities, which results in the nonreciprocal coupling. The ordinary cavity $a_{2}$ is driven by a classical laser field with frequency $\omega_{l}$. In the rotating frame with respect to the driving laser field $V=\exp[-i\omega_{l}t(a_{1}^{\dagger}a_{1}+a_{2}^{\dagger}a_{2})]$, the total Hamiltonian of the system is written as ($\hbar=1$)
\begin{eqnarray}\label{e01}
H&=&\Delta_{1}a_{1}^{\dagger}a_{1}+\Delta_{2}a_{2}^{\dagger}a_{2}+\omega_{m}b^{\dagger}b+\lambda_{1}a_{1}^{\dagger}a_{2}+\lambda_{2}a_{1}a_{2}^{\dagger}\cr\cr
&&-ga_{1}^{\dagger}a_{1}(b^{\dagger}+b)+Ea_{2}^{\dagger}+E^{\ast}a_{2},
\end{eqnarray}
where $\Delta_{j}=\omega_{j}-\omega_{l}$ ($j=1,2$) is the corresponding cavity-laser detuning. $\omega_{j}$ and $\omega_{m}$ are the resonance frequency of corresponding optical cavity and mechanical resonator. $\lambda_{j}$ represents the tunneling strength of photon jumping into cavity $a_{j}$. $|E|=\sqrt{2\kappa_{2}P/(\hbar\omega_{l})}$ is the amplitude of the driven laser field with power $P$ and $\kappa_{j}$ is the corresponding cavity decay rate~\cite{QST.4.024002}. In the mechanical displacement representation defined by a canonical transformation $V^{\prime}=\exp[g/\omega_{m}a_{1}^{\dagger}a_{1}(b^{\dagger}-b)]$, the nonlinear optomechanical coupling can be diagonalized as $g^{2}/\omega_{m}(a_{1}^{\dagger}a_{1})^{2}$. Under the condition of weak optomechanical coupling ($g/\omega_{m}\ll1$ in most actual optomechanical systems~\cite{RPP.83.026401}), the reduced Hamiltonian is given by (see Appendix~\ref{app1})
\begin{eqnarray}\label{e02}
H^{\prime}&=&\Delta_{1}a_{1}^{\dagger}a_{1}+\Delta_{2}a_{2}^{\dagger}a_{2}+\lambda_{1}a_{1}^{\dagger}a_{2}+\lambda_{2}a_{1}a_{2}^{\dagger}\cr\cr
&&-\chi(a_{1}^{\dagger}a_{1})^{2}+Ea_{2}^{\dagger}+E^{\ast}a_{2},
\end{eqnarray}
where $\chi=g^{2}/\omega_{m}$ is the Kerr-type nonlinear strength induced by the optomechanical coupling.

It should be pointed out that the classical laser field is applied to the ordinary cavity, not the optomechanical cavity~\cite{PhysRevA.99.043818}. This is because the perfect PB cannot occur when the optomechanical cavity is driven only by the classical laser field. The detailed discussion is given in the following text and Appendix~\ref{app2}.

\section{\label{sec.3} Photon statistic}
In this section, we respectively investigate the PB effect in the weak and strong coupling regions. These studies are analyzed in detail according to two different blockade mechanisms, and are carried out through analytically solving the non-Hermitian Schr\"odinger equation and numerically simulating the quantum master equation.

\subsection{\label{sec.3A}Analytical solution and numerical simulation}
Usually, the PB is characterized by the correlation function, which can be analytically obtained by solving the non-Hermitian Schr\"odinger equation. Here, the non-Hermitian Hamiltonian is directly given by adding phenomenologically the imaginary decay terms and reads
\begin{eqnarray}\label{e03}
H_{\mathrm{NM}}=H^{\prime}-i\frac{\kappa_{1}}{2}a_{1}^{\dagger}a_{1}-i\frac{\kappa_{2}}{2}a_{2}^{\dagger}a_{2}.
\end{eqnarray}
To obtain the analytical result of the correlation function, we calculate the Schr\"odinger equation under the weak driving condition ($E\ll\kappa_{2}$), i.e., $i\partial|\psi(t)\rangle/\partial t=H_{\mathrm{NM}}|\psi(t)\rangle$, where $|\psi(t)\rangle$ is the state of the system and the evolution space can be limited in the low-excitation subspace (up to 2). Thus the state of the system at any time can be expanded as
\begin{eqnarray}\label{e04}
|\psi(t)\rangle&=&\sum_{n_{1},n_{2}}^{n_{1}+n_{2}\leqslant 2}C_{n_{1}n_{2}}(t)|n_{1},n_{2}\rangle.
\end{eqnarray}
Here, $C_{n_{1}n_{2}}(t)$ is the probability amplitude of the state $|n_{1},n_{2}\rangle$, which represents $n_{1}$ photons being in the cavity $a_{1}$ and $n_{2}$ photons being in the cavity $a_{2}$. Under the weak driving condition, those probability amplitudes satisfy the relation of $|C_{n_{1}n_{2}}|_{n_{1}+n_{2}=2}\ll|C_{n_{1}n_{2}}|_{n_{1}+n_{2}=1}\ll|C_{00}|\simeq1$. Substituting the non-Hermitian Hamiltonian and the above state into the Schr\"odinger equation, we get a set of linear differential equations for the probability amplitudes, which reads
\begin{eqnarray}\label{e05}
i\frac{\partial C_{10}}{\partial t}&=&\left(\Delta_{1}-i\frac{\kappa_{1}}{2}-\chi\right)C_{10}+\lambda_{1}C_{01}+E^{\ast}C_{11},\cr\cr
i\frac{\partial C_{01}}{\partial t}&=&\left(\Delta_{2}-i\frac{\kappa_{2}}{2}\right)C_{01}+\lambda_{2}C_{10}+EC_{00}+\sqrt{2}E^{\ast}C_{02},\cr\cr
i\frac{\partial C_{20}}{\partial t}&=&\left(2\Delta_{1}-i\kappa_{1}-4\chi\right)C_{20}+\sqrt{2}\lambda_{1}C_{11},\cr\cr
i\frac{\partial C_{11}}{\partial t}&=&\left(\Delta_{1}-i\frac{\kappa_{1}}{2}+\Delta_{2}-i\frac{\kappa_{2}}{2}-\chi\right)C_{11}\cr\cr
&&+\sqrt{2}\left(\lambda_{1}C_{02}+\lambda_{2}C_{20}\right)+EC_{10},\cr\cr
i\frac{\partial C_{02}}{\partial t}&=&\left(2\Delta_{2}-i\kappa_{2}\right)C_{02}+\sqrt{2}\lambda_{2}C_{11}+\sqrt{2}EC_{01}.
\end{eqnarray}

The above equations can be solved directly to obtain the dynamical result of state evolution. However, because what we are interested in is the steady-state photon statistical property, we thus just give the steady-state result, which can be simplified through some approximate processes, e.g., ignoring those higher-order terms, $E^{\ast}=E$, $\Delta_{1}=\Delta_{2}$, and $\kappa_{1}=\kappa_{2}$. The approximate steady-state solution is given as
\begin{eqnarray}\label{e06}
C_{10}&=&-4E\lambda_{1}/M,\cr\cr
C_{01}&=&2E\left(2\Delta_{2}-2\chi-i\kappa_{2}\right)/M,\cr\cr
C_{20}&=&8\sqrt{2}E^{2}\lambda_{1}^{2}\left(\chi-2\Delta_{2}+i\kappa_{2}\right)/N,\cr\cr
C_{11}&=&8E^{2}\lambda_{1}\left(\chi-2\Delta_{2}+i\kappa_{2}\right)\left(4\chi-2\Delta_{2}+i\kappa_{2}\right)/N,\cr\cr
C_{02}&=&2\sqrt{2}E^{2}\big[4\lambda_{1}\lambda_{2}\chi+\left(\chi-2\Delta_{2}+i\kappa_{2}\right)\cr\cr
&&\times\left(2\chi-2\Delta_{2}+i\kappa_{2}\right)\left(4\chi-2\Delta_{2}+i\kappa_{2}\right)\big]/N,
\end{eqnarray}
with
\begin{eqnarray}\label{e07}
M&=&\big[4\lambda_{1}\lambda_{2}+\left(2\Delta_{2}-i\kappa_{2}\right)\left(2\chi-2\Delta_{2}+i\kappa_{2}\right)\big],\cr\cr
N&=&\big[4\lambda_{1}\lambda_{2}(2\chi-2\Delta_{2}+i\kappa_{2})+(2\Delta_{2}-i\kappa_{2})\cr\cr
&&\times(\chi-2\Delta_{2}+i\kappa_{2})(4\chi-2\Delta_{2}+i\kappa_{2})\big]M.
\end{eqnarray}

By combining the above steady-state analytical result with the definition of equal-time second-order correlation function $g_{j}^{(2)}(0)=\langle a_{j}^{\dagger}a_{j}^{\dagger}a_{j}a_{j}\rangle/\langle a_{j}^{\dagger}a_{j}\rangle^{2}$, the final analytical result reads
\begin{eqnarray}\label{e08}
g_{1}^{(2)}(0)&=&\frac{\langle a_{1}^{\dagger}a_{1}^{\dagger}a_{1}a_{1}\rangle}{\langle a_{1}^{\dagger}a_{1}\rangle^{2}}\simeq\frac{2|C_{20}|^{2}}{|C_{10}|^{4}},\cr\cr
g_{2}^{(2)}(0)&=&\frac{\langle a_{2}^{\dagger}a_{2}^{\dagger}a_{2}a_{2}\rangle}{\langle a_{2}^{\dagger}a_{2}\rangle^{2}}\simeq\frac{2|C_{02}|^{2}}{|C_{01}|^{4}},
\end{eqnarray}
which characterizes the probability of observing two photons being in the $i$th cavity at the same time. Here, $g_{j}^{(2)}(0)>1$ represents the photon bunching effect, otherwise it represents the photon anti-bunching effect and satisfies the sub-Poissonian light statistics. Specifically, the perfect PB effect can be characterized by a necessary condition $g_{j}^{(2)}(0)=0$, namely, $|C_{20}|=0$ or $|C_{02}|=0$. It is easy to find that only $|C_{02}|=0$ can be satisfied with the following parameter relations:
\begin{eqnarray}\label{e09}
\Delta_{2}^{\pm}&=&\pm\frac{\sqrt{3\kappa_{2}^2+7\chi^2}}{6}+\frac{7\chi}{6},\cr\cr
\lambda_{2}^{\pm}&=&\mp\frac{\sqrt{3\kappa_{2}^2+7\chi^2}}{54\lambda_{1}}\left(\frac{12\kappa_{2}^{2}}{\chi}+7\chi\right)-\frac{5\chi^{2}}{27\lambda_{1}},
\end{eqnarray}
where we obtain two different optimal relations distinguished by `$\pm$'. When the above relations are satisfied, the two-photon excited state $|0,2\rangle$ can be completely suppressed due to the ideal destructive quantum interference between two different excitation paths ($|0,0\rangle\rightarrow|0,1\rangle\rightarrow|0,2\rangle$ and $|0,0\rangle\rightarrow|0,1\rangle\xrightarrow{\lambda_{1}}|1,0\rangle\rightarrow|1,1\rangle\xrightarrow{\lambda_{2}}|0,2\rangle$). So we focus on studying the photon statistic in the driven cavity in the following parts. However, it is worth noting that the study does not mean that no PB occurs in the optomechanical cavity, because it only needs to satisfy $g_{1}^{(2)}(0)\ll1$.

The above calculations have given the analytical result of the equal-time second-order correlation function through approximate processes. To verify the analytical result, we give the exact numerical simulation by utilizing the quantum master equation
\begin{eqnarray}\label{e10}
\frac{\partial\rho}{\partial t}&=&-i\left[H,\rho\right]+\kappa_{1}\mathcal{L}[a_{1}]\rho+\kappa_{2}\mathcal{L}[a_{2}]\rho\cr\cr
&&+\gamma_{m}(n_{\mathrm{th}}+1)\mathcal{L}[b]\rho+\gamma_{m}n_{\mathrm{th}}\mathcal{L}[b^{\dagger}]\rho.
\end{eqnarray}
where $\mathcal{L}[o]\rho=o\rho o^{\dagger}-(o^{\dagger}o\rho+\rho o^{\dagger}o)$ represents the Lindblad operator for the arbitrary system operator $o$. $\gamma_{m}$ is the mechanical damping rate and $n_{\mathrm{th}}$ is the mean thermal phonon number. $\rho$ is the dynamical density matrix of the system, which can reach its own steady state $\rho_{s}$ after a long evolution time. Meanwhile, the numerical result of the equal-time second-order correlation function is
\begin{eqnarray}\label{e11}
g_{2}^{(2)}(0)=\frac{\mathrm{Tr}(a_{2}^{\dagger}a_{2}^{\dagger}a_{2}a_{2}\rho_{s})}{[\mathrm{Tr}(a_{2}^{\dagger}a_{2}\rho_{s})]^{2}}.
\end{eqnarray}

The analytical and numerical results of the equal-time second-order correlation function have been respectively given through the above calculations. Furthermore, the PB is usually characterized through the delayed second-order correlation function, which represents the probability of detecting another photon after a finite-time delay $\tau$. Generally, the delayed second-order correlation function at any time is defined by
\begin{eqnarray}\label{e12}
g_{2}^{(2)}(\tau)=\frac{\langle a_{2}^{\dagger}(t)a_{2}^{\dagger}(t+\tau)a_{2}(t+\tau)a_{2}(t)\rangle}{\langle a_{2}^{\dagger}(t)a_{2}(t)\rangle\langle a_{2}^{\dagger}(t+\tau)a_{2}(t+\tau)\rangle},
\end{eqnarray}
which is not convenient to solve directly. To calculate the delayed second-order correlation function in the Schr\"odinger picture, the above definition needs a transformation, i.e., $\langle o(t)\rangle=\mathrm{Tr}[oU(t)\rho U(t)^{\dagger}]$, where $U(t)$ represents the evolution operator of the system. When the system reaches its steady state, the delayed second-order correlation function in the Schr\"odinger picture is rewritten as
\begin{eqnarray}\label{e13}
g_{2}^{(2)}(\tau)=\frac{\mathrm{Tr}[a_{2}^{\dagger}a_{2}U(\tau)a_{2}\rho_{s}a_{2}^{\dagger}U^{\dagger}(\tau)]}{[\mathrm{Tr}(a_{2}^{\dagger}a_{2}\rho_{s})]^{2}},
\end{eqnarray}
which is convenient to solve through the above quantum master equation~(\ref{e10}). Now, we have given the analytical and numerical results of the correlation function, and discussed the calculation method of the delayed second-order correlation functions. Next, we study the photon statistic and emission properties of the system and analyze their physical essence based on different blockade mechanisms.

\subsection{\label{sec.3B}Weak coupling}
In the usual proposals, the required coupling is very strong to structure the destructive quantum interference~\cite{PhysRevA.96.053810} or the anharmonic eigenenergy spectrum~\cite{PhysRevLett.107.063601,PhysRevLett.107.063602}, namely, $\{\lambda_{1},\lambda_{2}\}>\kappa_{2}$ and $g\sim\omega_{m}$. Here, we strictly analyze the optimal parameter relation via the UPB mechanism and find that the strong coupling is unnecessary in our proposal.

\begin{figure}
	\includegraphics[width=0.49\linewidth]{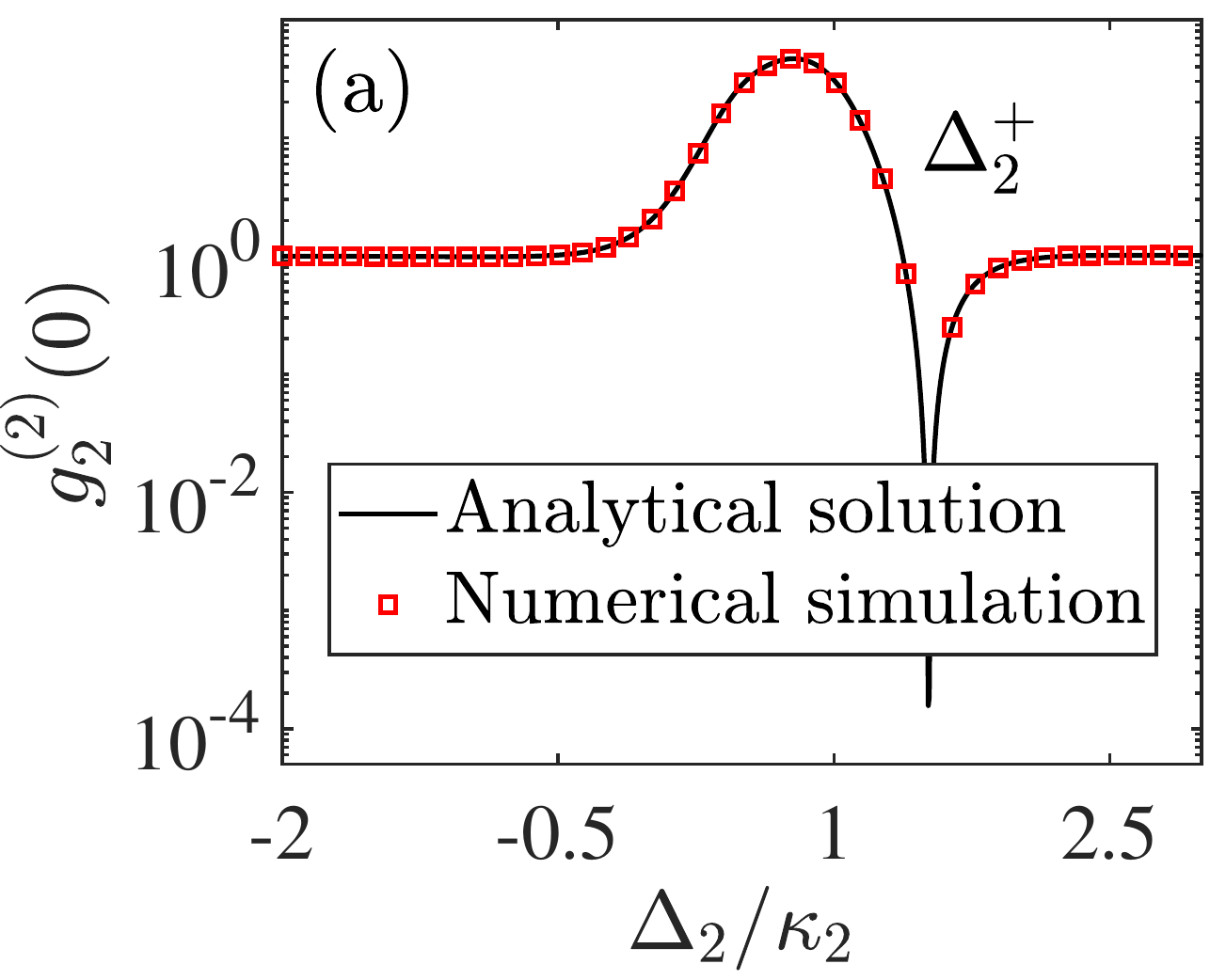}
	\hspace{0in}%
	\includegraphics[width=0.49\linewidth]{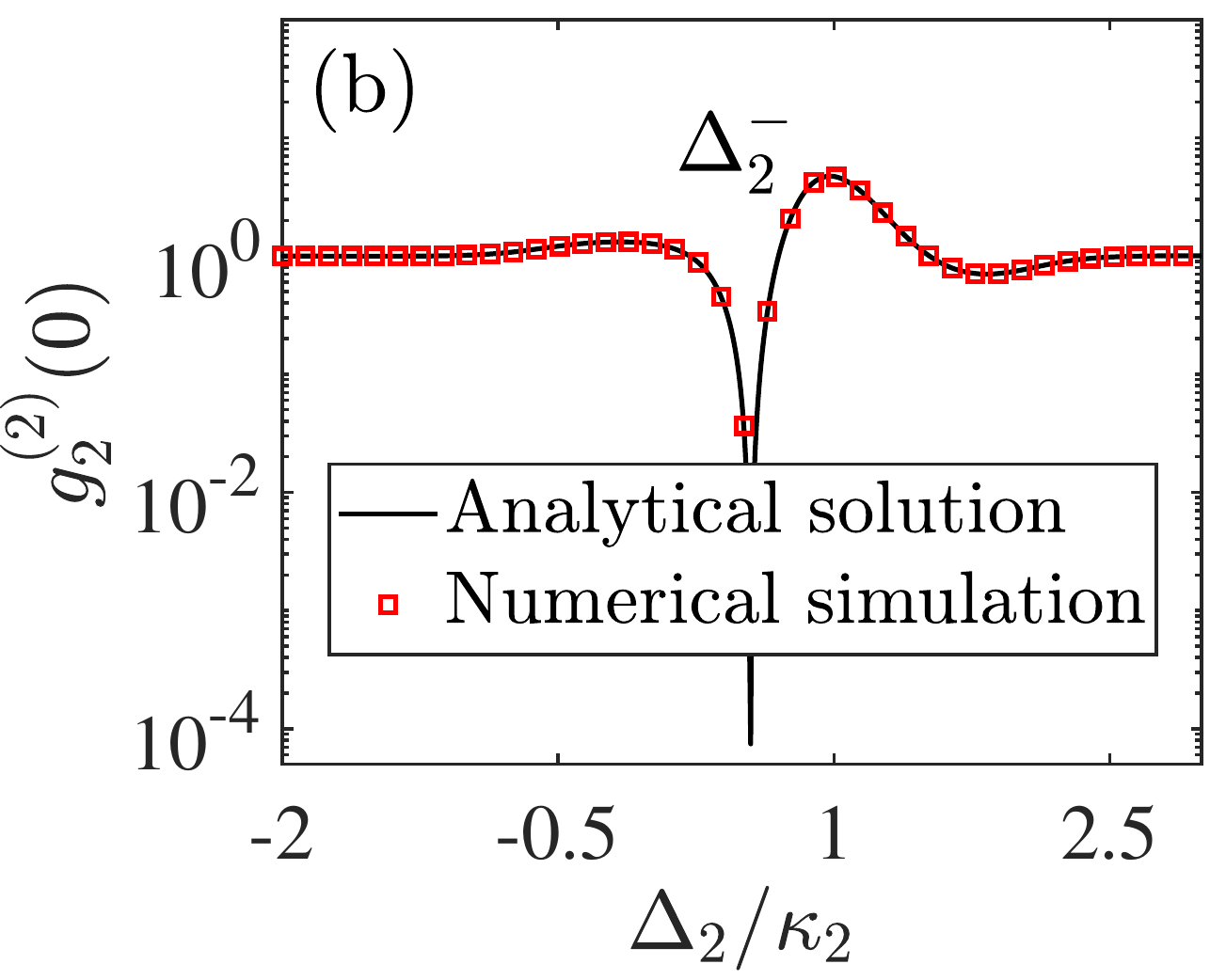}
	\vspace{0in}%
	\includegraphics[width=0.49\linewidth]{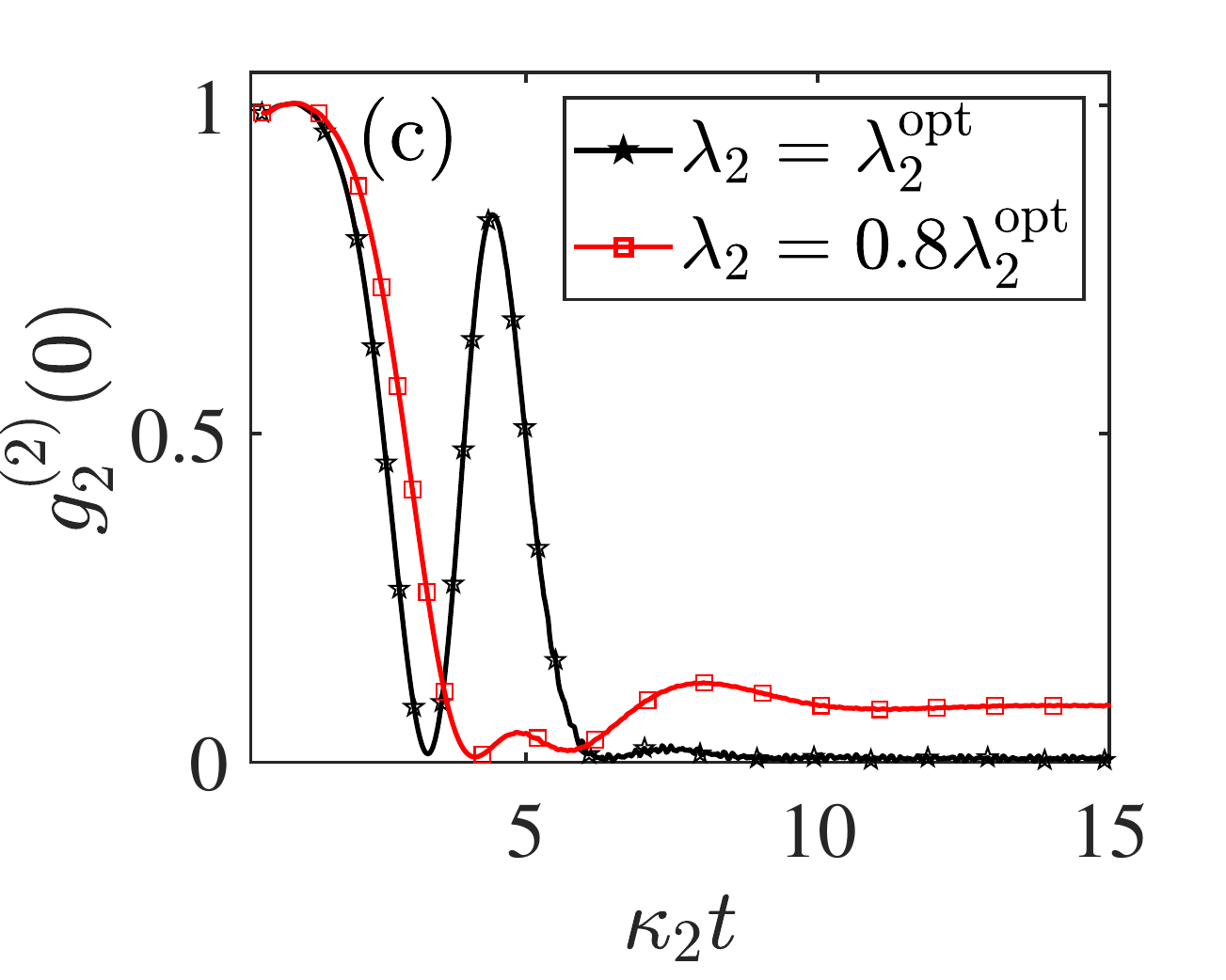}
	\hspace{0in}%
	\includegraphics[width=0.49\linewidth]{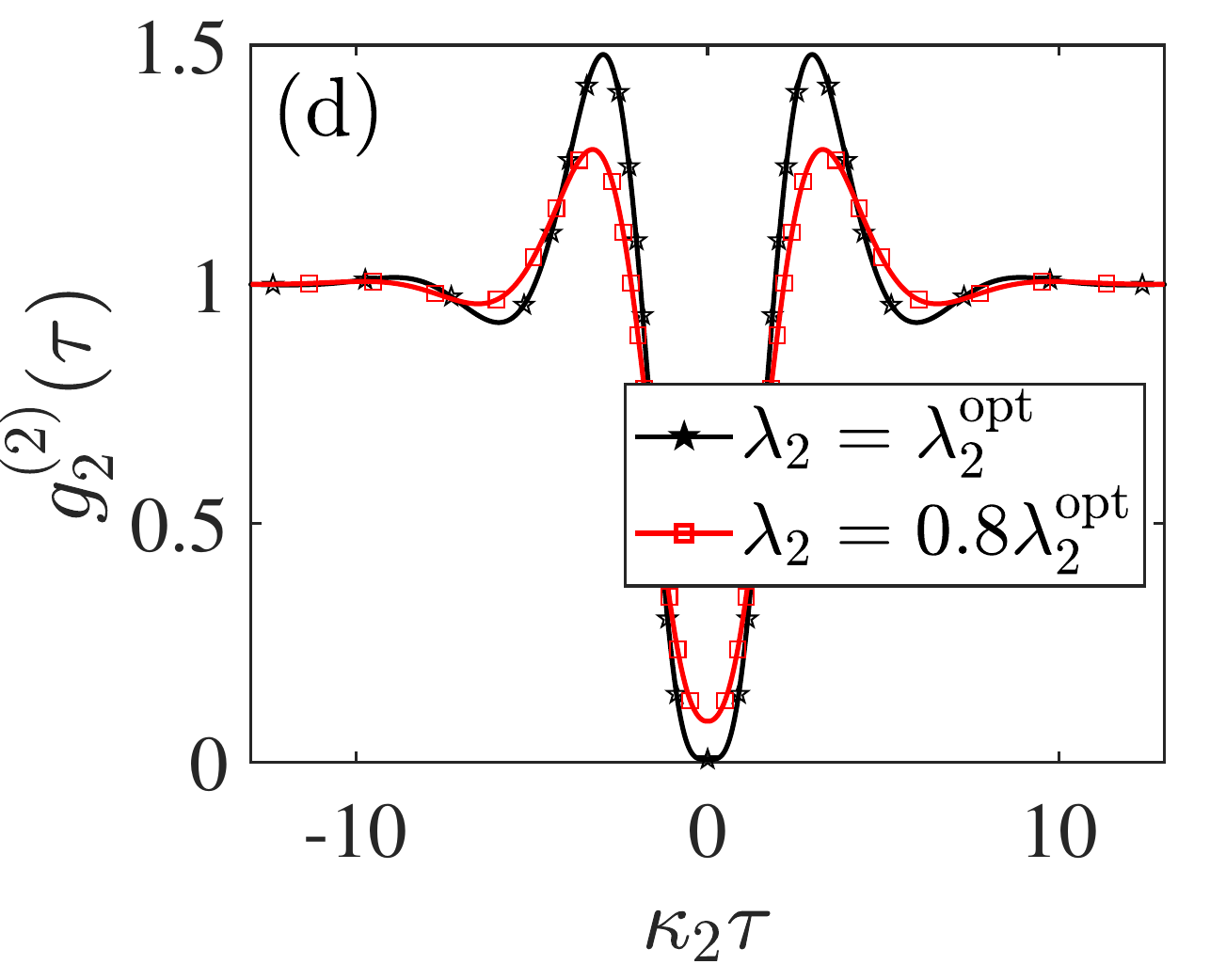}
	\caption{\label{fig:2}(a) and (b) represent the equal-time second-order correlation function versus the cavity-laser detuning with the optimal parameter relations designated by `$+$' and `$-$' in Eq.~(\ref{e09}), respectively. (c) The dynamical evolution of the equal-time second-order correlation function with the optimal parameter $\Delta_{2}^{-}$ and $\lambda_{2}^{-}$. Here, we have assumed the system initial state is in the vacuum state. (d) The delayed second-order correlation function versus the time-delay. The system parameters are set as $\kappa_{2}=2\pi\times0.15~\mathrm{MHz}$, $E/\kappa_{2}=0.02$, $\lambda_{1}/\kappa_{2}=0.95$, $\omega_{m}=2\pi\times75~\mathrm{MHz}$, $\gamma_{m}=\omega_{m}/10^{6}$, $n_{\mathrm{th}}=0$, and $g/\omega_{m}=0.042$.}
\end{figure}

To verify our analysis, we plot the equal-time second-order correlation function versus the cavity-laser detuning $\Delta_{2}$ through analytical and numerical methods in Figs.~\ref{fig:2}(a) and \ref{fig:2}(b), respectively, where the analytical and numerical results agree well with each other and the system parameters satisfy $\{\lambda_{1},\lambda_{2}\}<\kappa_{2}$ and $g/\omega_{m}\ll1$. We can see that the perfect PB occurs with both the optimal parameters in Eq.~(\ref{e09}). After that, we only take the optimal relation designated by `$-$' as an example. The dynamical evolution of the correlation function is shown in Fig.~\ref{fig:2}(c), which proves the system reaches a steady state after a long evolution time and the perfect PB occurs with the optimal system parameters. As a comparison, when the nonreciprocal coupling does not take the optimal value $\lambda_{2}^{\mathrm{opt}}$, the correlation function cannot reach 0 when the system reaches its steady state. That is because the non-optimal coupling does not achieve the complete destructive quantum interference so that the two-photon excitation occurs. Finally, we numerically calculate the delayed second-order correlation function according to Eq.~(\ref{e13}) and show the result in Fig.~\ref{fig:2}(d). Similarly, when the system parameters are not optimal, the delayed second-order correlation function is also calculated and shown. The results indicate that the delayed second-order correlation function is always larger than the equal-time one, i.e., $g_{2}^{(2)}(\tau)>g_{2}^{(2)}(0)$, which characterizes the non-classical anti-bunching effect of photons.

\begin{figure}
	\includegraphics[width=0.49\linewidth]{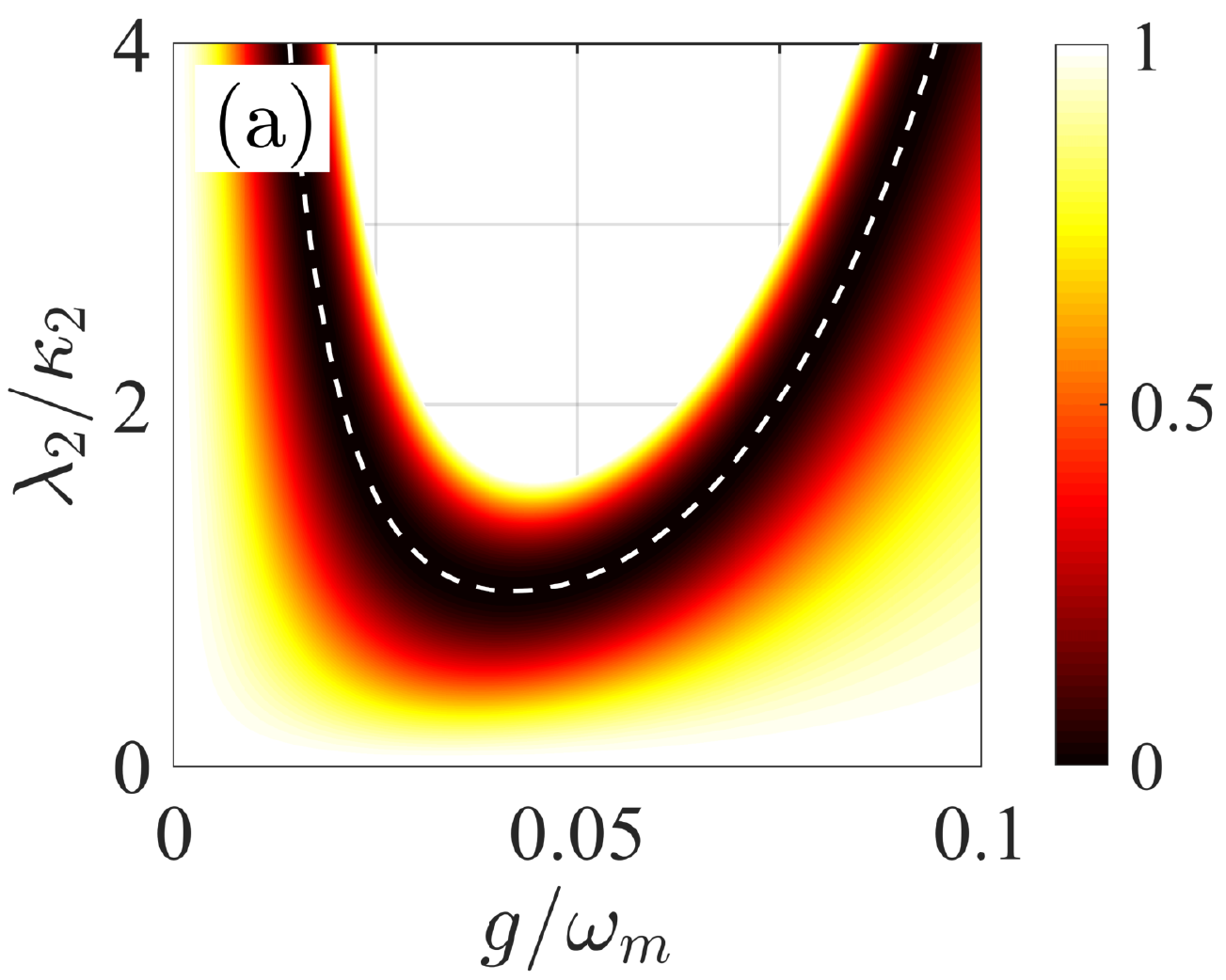}
	\hspace{0in}%
	\includegraphics[width=0.49\linewidth]{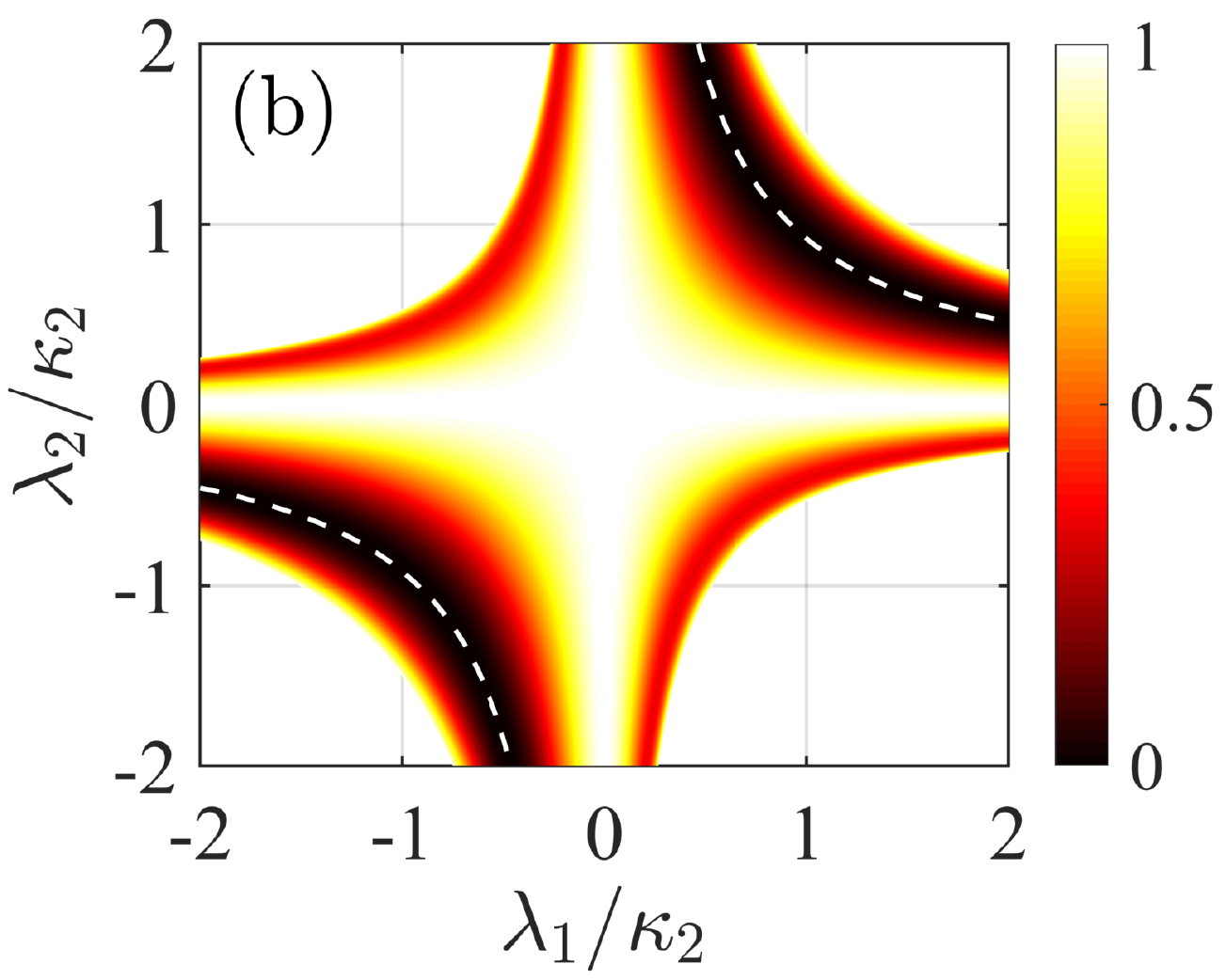}
	\vspace{0in}%
	\includegraphics[width=0.49\linewidth]{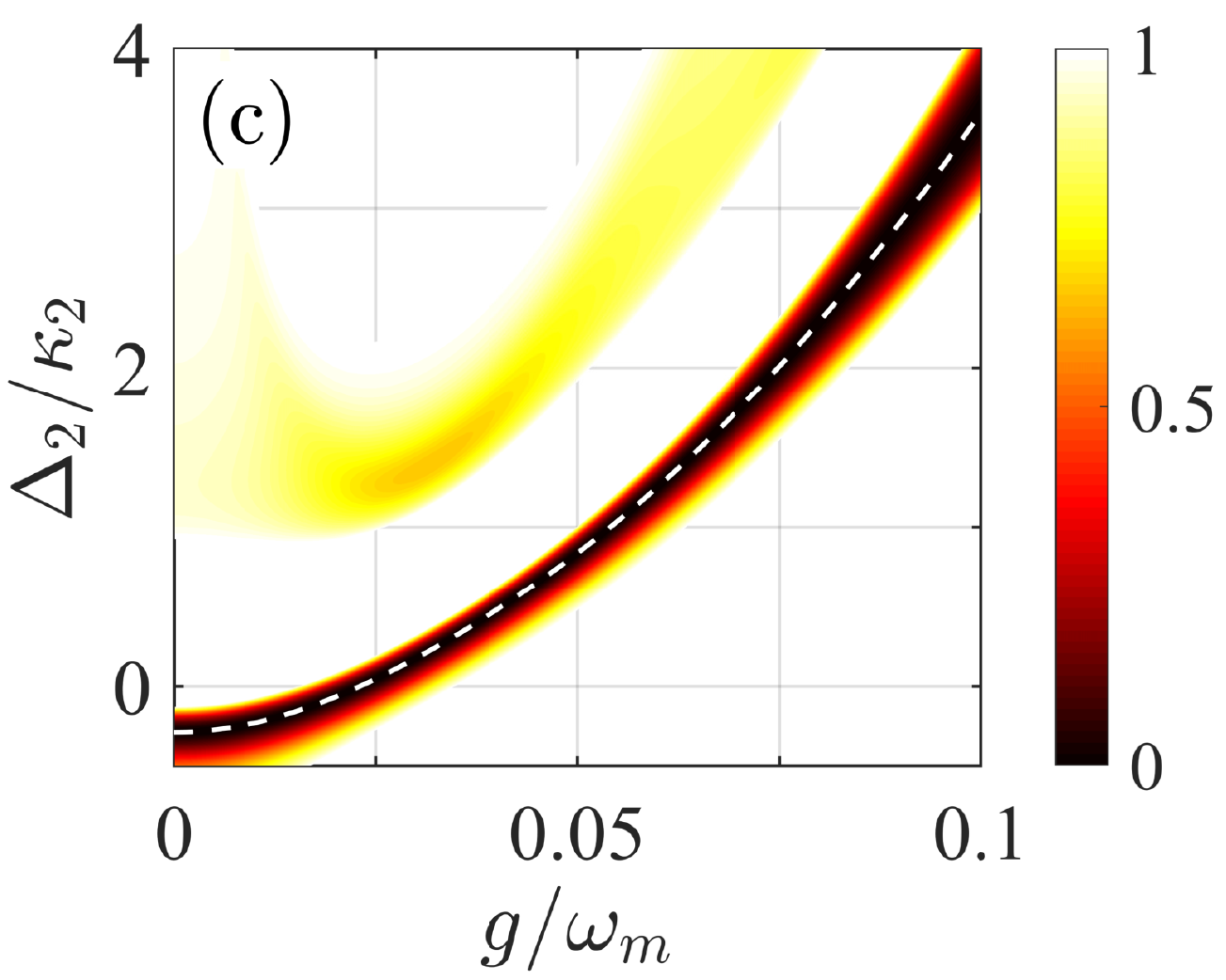}
	\hspace{0in}%
	\includegraphics[width=0.49\linewidth]{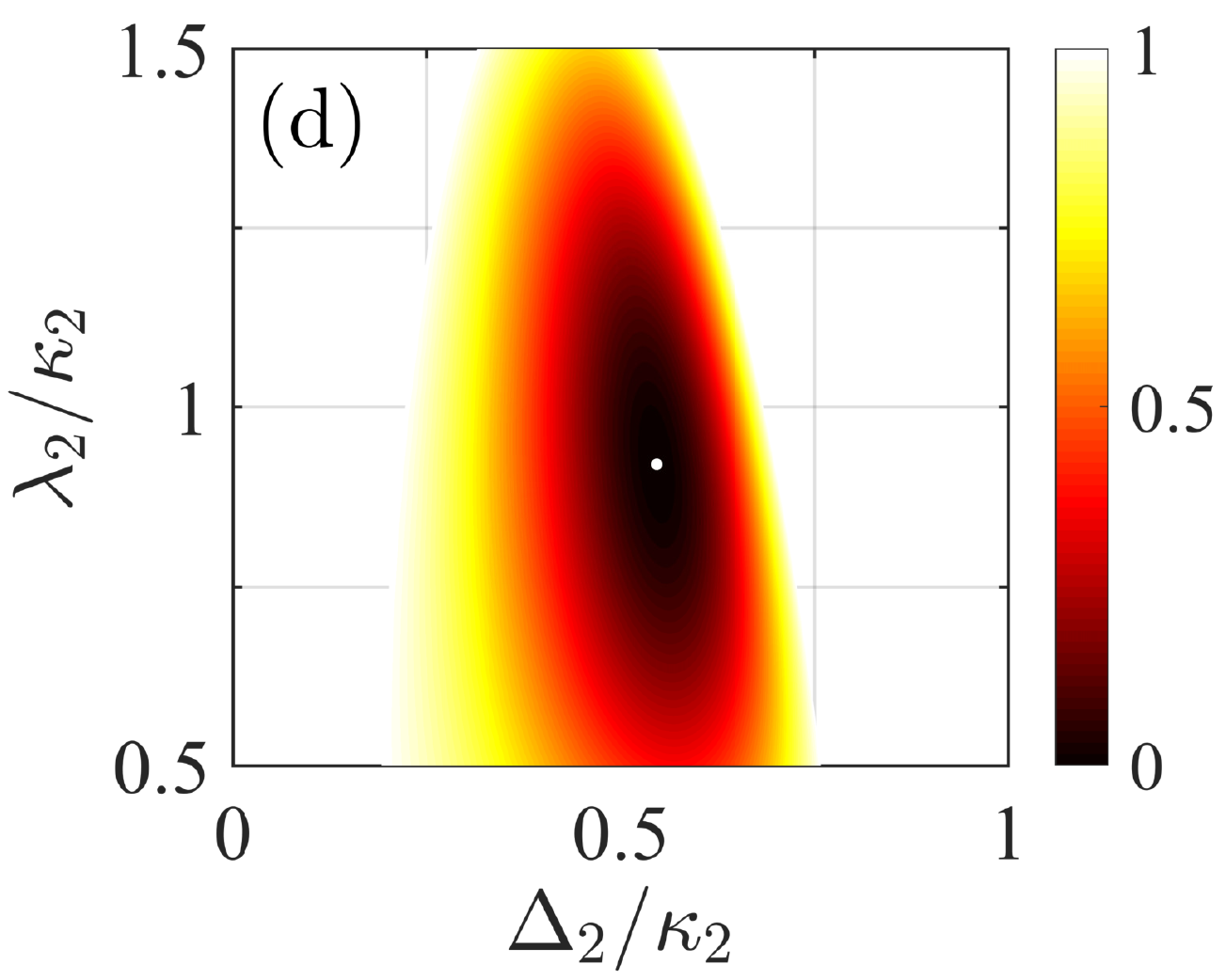}
	\caption{\label{fig:3}The equal-time second-order correlation function versus different system parameters. Here, the dashed white line and white point represent the optimal parameter relation. The blank regions represent the equal-time second-order correlation function is larger than 1 and the photon occurs bunching effect.}	
\end{figure}

Moreover, we also study the effect of parameter fluctuations on the equal-time second-order correlation function and show those results in Fig.~\ref{fig:3}. To further clarify whether the perfect PB can be achieved for $\{\lambda_{1},\lambda_{2}\}<\kappa_{2}$ and $g/\omega_{m}\ll1$, we analyze the optimal parameter relation and find that the required optimal nonreciprocal coupling is related to the optomechanical coupling strength. The minimal value of the optimal nonreciprocal coupling occurs at $g/\omega_{m}=0.042$, which can be derived from Eq.~(\ref{e09}) and seen from Fig.~\ref{fig:3}(a). And when the optomechanical coupling strength is $g/\omega_{m}=0.042$, the optimal nonreciprocal coupling satisfies an inverse proportional function, which reads $\lambda_{1}\lambda_{2}/\kappa_{2}^{2}=0.92$, as shown in Fig.~\ref{fig:3}(b). That means the perfect PB can occur even with $\{\lambda_{1},\lambda_{2}\}<\kappa_{2}$ and $g/\omega_{m}\ll1$, which breaks the strong-coupling limit of UPB mechanism. In addition, the location of perfect blockade occurring is also related to the optomechanical coupling, as shown in Fig.~\ref{fig:3}(c). That is to say, the optomechanical coupling is a core role for the perfect blockade, which determines the location of blockade and the optimal nonreciprocal coupling strength in our scheme, as shown in Fig.~\ref{fig:3}(d). Contrarily, whether the optomechanical coupling is vanishing or too large, the required nonreciprocal coupling strength is larger than the cavity decay, which means that the perfect photon cannot be achieved under the weak coupling region.

\begin{figure}
	\includegraphics[width=0.7\linewidth]{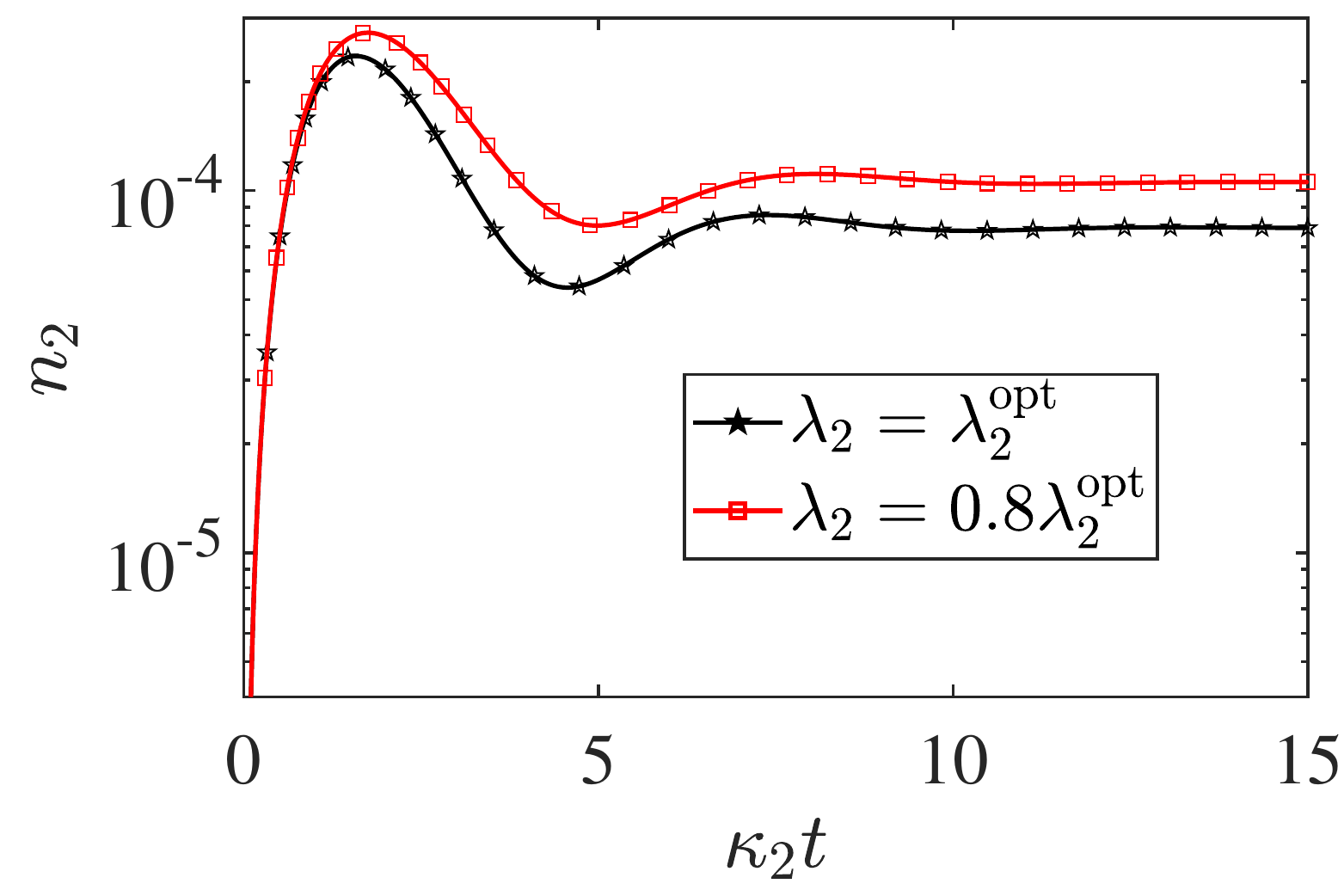}
	\caption{\label{fig:4}The dynamical evolution of the intracavity photon number.}	
\end{figure}

To measure the efficiency of single-photon emission, we discuss the dynamical evolution of the photon number ($n_{2}=\langle a_{2}^{\dagger}a_{2}\rangle$) with different nonreciprocal coupling strengths, as shown in Fig.~\ref{fig:4}. We can see that the intracavity photon number reaches a steady-state value $n_{2}\sim10^{-4}$ when the system is stable. Meanwhile, the efficiency of single-photon emission is estimated by the output photon flux $\kappa_{2}\langle a_{2}^{\dagger}a_{2}\rangle$. Furthermore, one can see from Figs.~\ref{fig:2}(c) and \ref{fig:4} that, although the efficiency of single-photon emission is improved with an imperfect nonreciprocal coupling $0.8\lambda_{2}^{\mathrm{opt}}$, the quality of the single-photon source is declined.

Based on the above analysis and discussion, we find that the perfect PB can be obtained even with the weak coupling parameters in the nonreciprocal coupling system, i.e., $\{\lambda_{1},\lambda_{2}\}<\kappa_{2}$ and $g/\omega_{m}\ll1$. The obtained PB belongs to the UPB mechanism so that the derived optimal parameter relation is also the unconventional optimal parameter relation. Next, we study the different blockade mechanisms with the strong coupling parameters and discuss their influence on blockade effect.

\subsection{\label{sec.3C}Strong coupling}
Different from the weak coupling, the CPB phenomenon would become obvious because the energy-level splitting induced by the strong coupling causes a large transition detuning between single-photon and two-photon excitations. So we discuss both conventional and unconventional blockade mechanisms under the strong coupling region. For the unconventional blockade, the above calculation results are still applicable. Therefore, we analyze the conventional blockade according to the energy spectrum of different excitations. Expanding the system Hamiltonian without the laser driving term, i.e., $H_{\mathrm{ND}}=\Delta_{1}a_{1}^{\dagger}a_{1}+\Delta_{2}a_{2}^{\dagger}a_{2}-\chi(a_{1}^{\dagger}a_{1})^{2}+\lambda_{1}a_{1}^{\dagger}a_{2}+\lambda_{2}a_{1}a_{2}^{\dagger}$, we directly solve the eigenvalue equation of the system to obtain the energy spectrum in different excitation subspaces. For the sake of simplicity, we still assume $\Delta_{1}=\Delta_{2}$ and $\kappa_{1}=\kappa_{2}$ in subsequent calculations. Meanwhile, it is easy to derive the system eigenvalues in different excitation subspaces, e.g., $\varepsilon_{0}=0$ for the zero excitation and $\varepsilon_{1\pm}=\pm\sqrt{\lambda_{1}\lambda_{2}+\chi^{2}/4}-\chi/2+\Delta_{2}$ for the single excitation. However, the eigenvalues of two excitations are too cumbersome to show here. Utilizing the theory of conventional blockade mechanism (resonant transition between the zero and single excitations), we can derive the locations of CPB occurring
\begin{eqnarray}\label{e14}
\Delta_{2}=\pm\sqrt{\lambda_{1}\lambda_{2}+\frac{\chi^{2}}{4}}+\frac{\chi}{2}.
\end{eqnarray}
Thus, when the nonreciprocal coupling takes the above unconventional optimal parameter relation, there will be three locations occurring PB, where two conventional and one unconventional blockades are located at $\pm\sqrt{\lambda_{1}\lambda_{2}+\chi^{2}/4}+\chi/2$ and $(7\chi-\sqrt{3\kappa_{2}^2+7\chi^2})/6$, respectively.

\begin{figure}
	\includegraphics[width=0.49\linewidth]{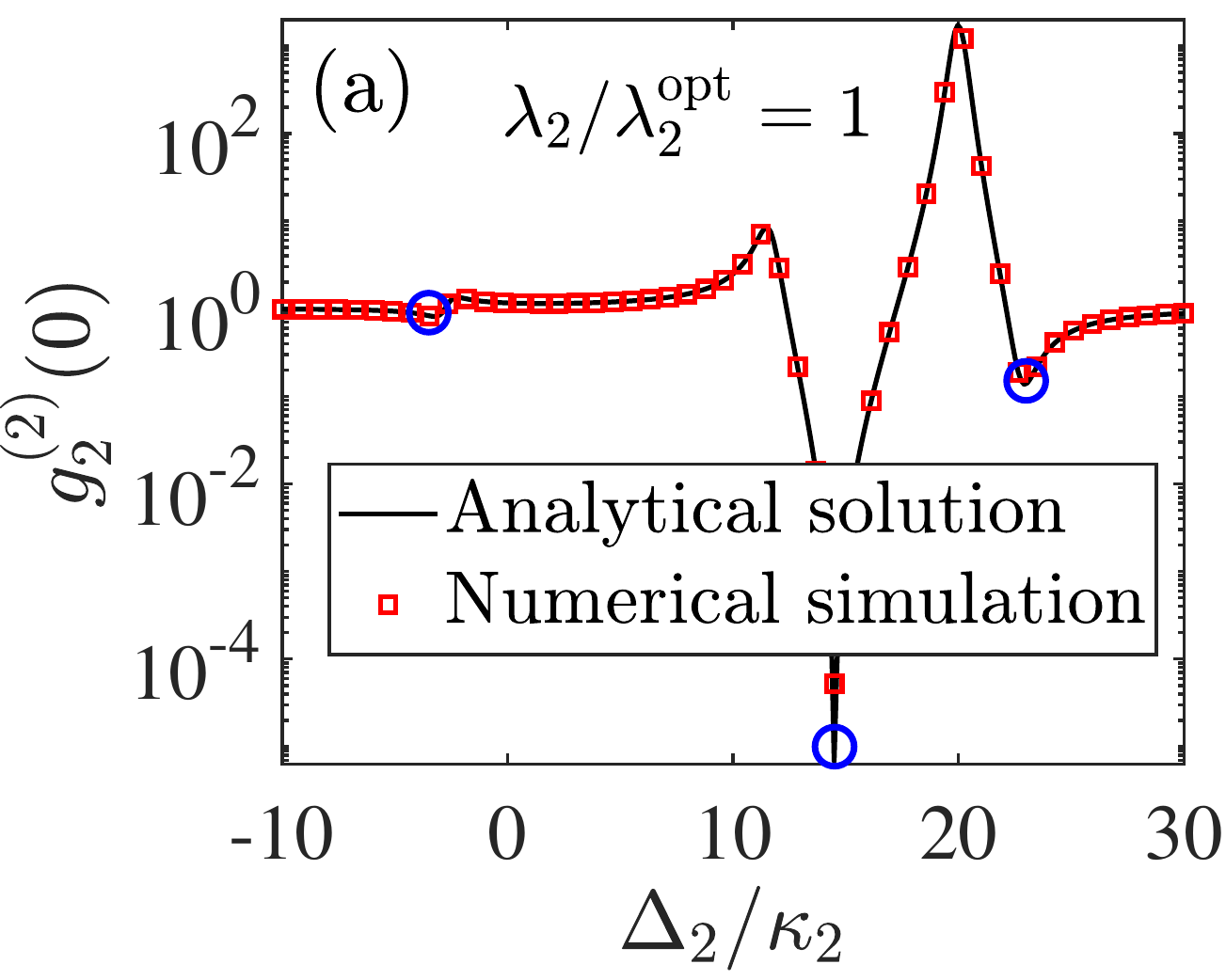}
	\hspace{0in}%
	\includegraphics[width=0.49\linewidth]{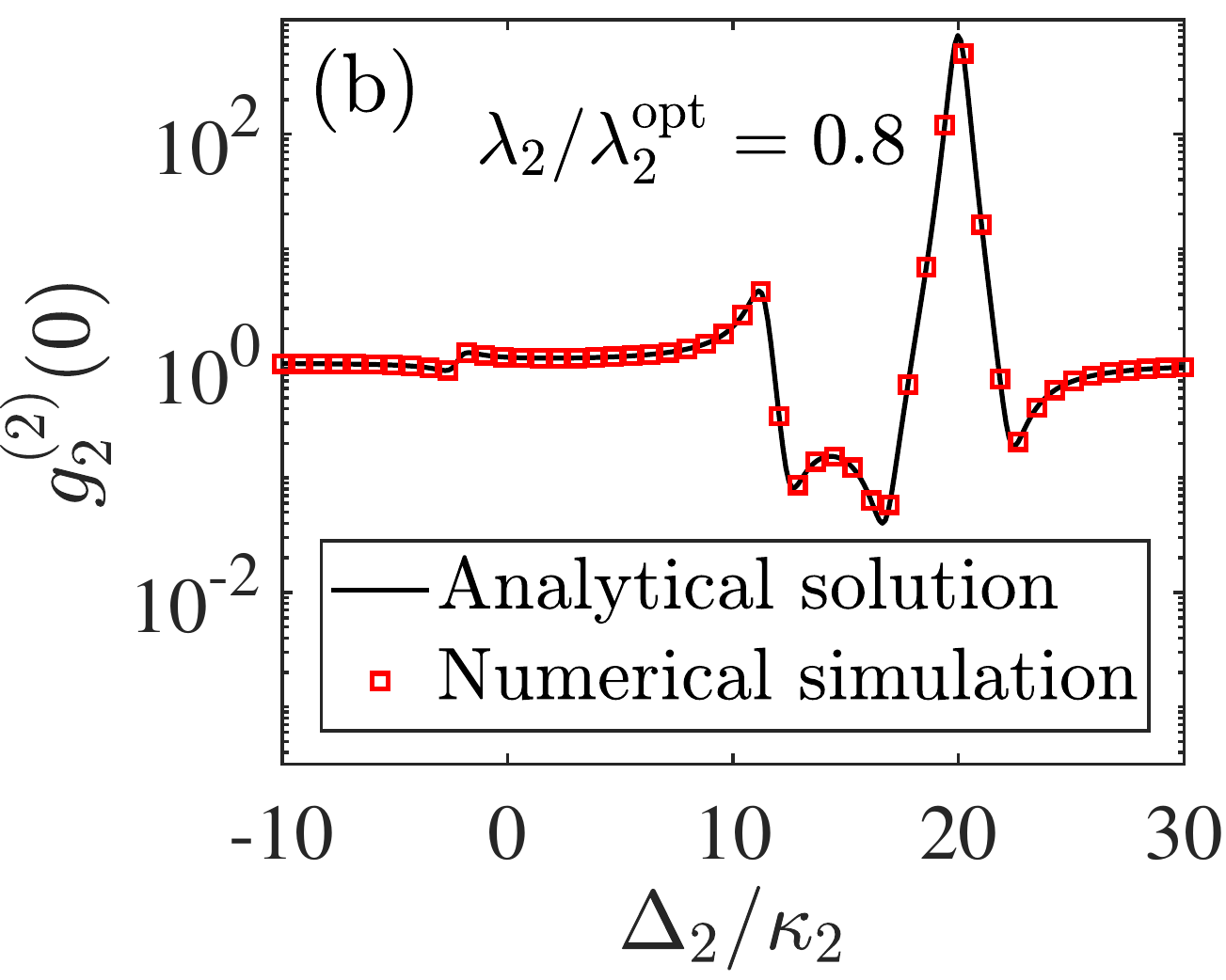}
	\vspace{0in}%
	\includegraphics[width=0.49\linewidth]{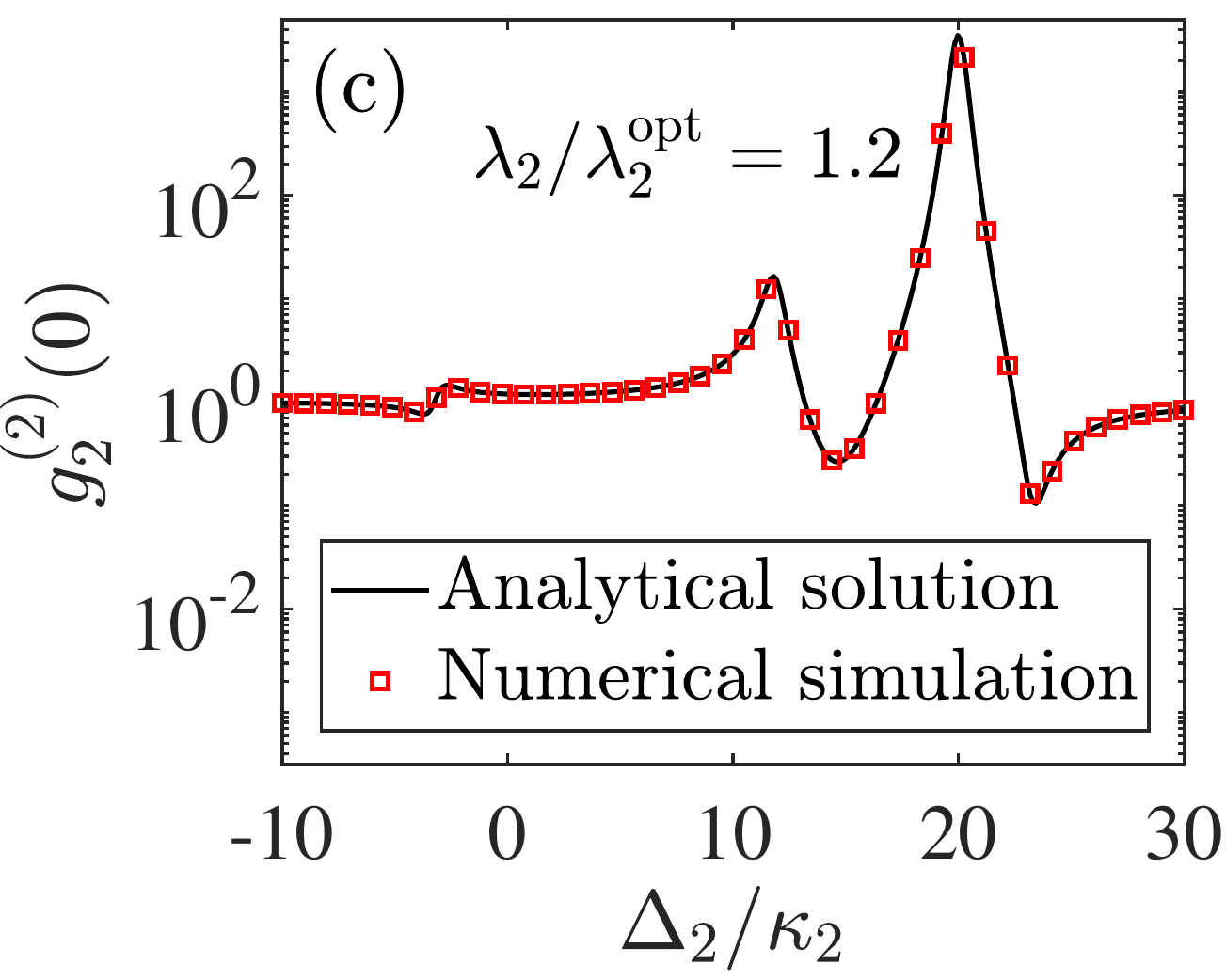}
	\hspace{0in}%
	\includegraphics[width=0.49\linewidth]{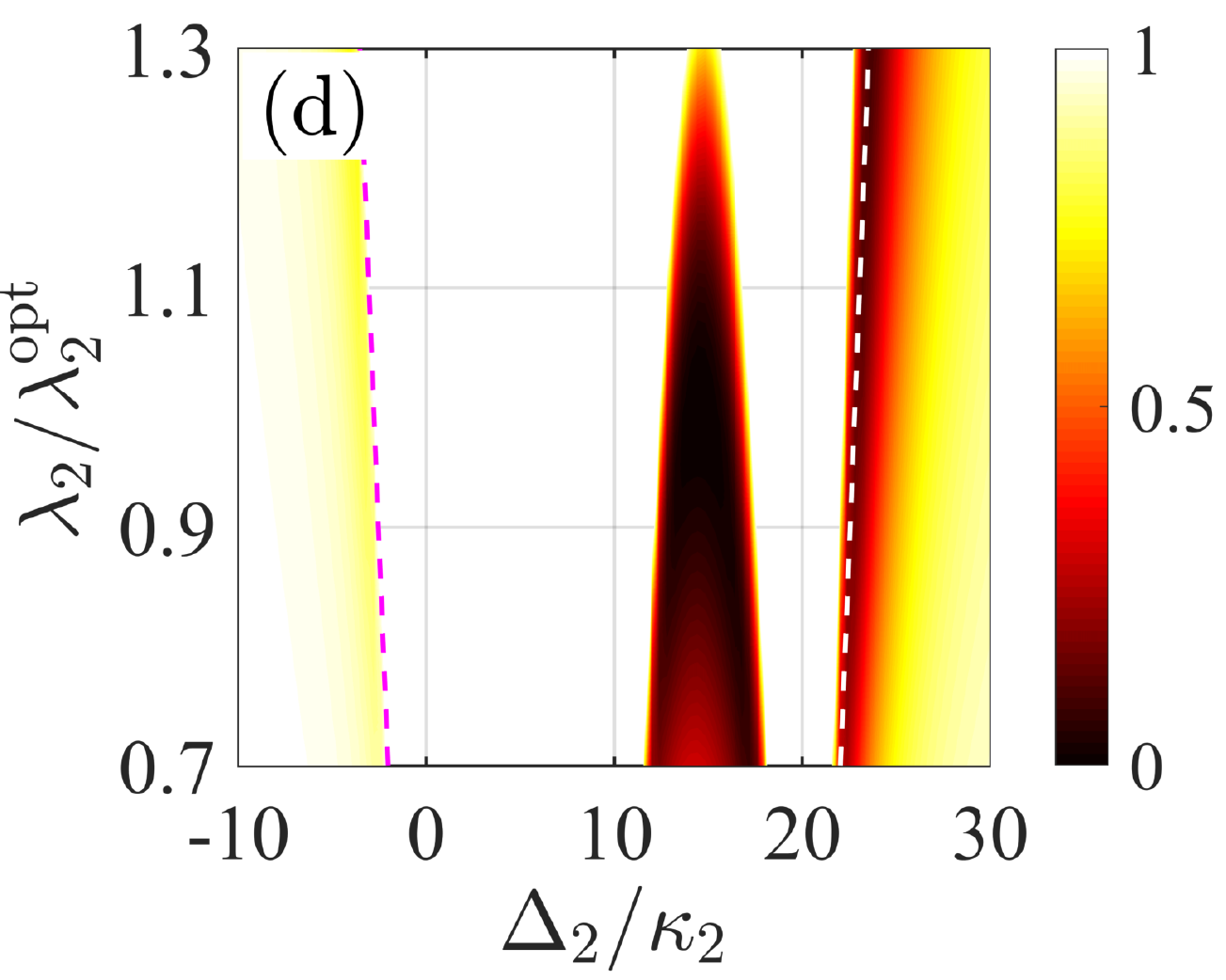}
	\caption{\label{fig:5}The equal-time second-order correlation function in the strong coupling region, e.g., $g/\omega_{m}=0.2$ and $\lambda_{1}/\kappa_{2}=8$. (a) $\lambda_{2}=\lambda_{2}^{\mathrm{opt}}$; (b) $\lambda_{2}=0.8\lambda_{2}^{\mathrm{opt}}$; (c) $\lambda_{2}=1.2\lambda_{2}^{\mathrm{opt}}$; (d) The correlation function versus both the detuning and the nonreciprocal coupling. Here, the dashed white and magenta lines represent the conventional optimal parameter relation of the single-excitation resonant $\Delta_{2}=\pm\sqrt{\lambda_{1}\lambda_{2}+\chi^{2}/4}+\chi/2$.}	
\end{figure}

To verify the above analysis, we plot the analytical and numerical results of the equal-time second-order correlation function in Fig.~\ref{fig:5}(a). For the strong parameters, e.g., $g/\omega_{m}=0.2$ and $\lambda_{1}/\kappa_{2}=8$, the PB locations occur at the locations $\Delta_{2}/\kappa_{2}=-3.5$, $14.5$, and $23.5$ [see the three blue circles in Fig.~\ref{fig:5}(a)], which respectively correspond to the different blockade mechanism. Specifically, the two dips located at both sides belong to the CPB and the middle one is the location of the UPB occurring. Furthermore, to study the effect of nonreciprocal coupling on different blockade mechanisms, we also calculate the correlation functions with the non-optimal nonreciprocal coupling and show them in Figs.~\ref{fig:5}(b)-\ref{fig:5}(d). For the UPB, when the nonreciprocal coupling is smaller than the optimal value, the the UPB splits into two dips, as shown in Fig.~\ref{fig:5}(b). As a contrast, the blockade location of the UPB does not change but the blockade effect is worse with a larger nonreciprocal coupling [see Fig.~\ref{fig:5}(c)]. That is because the amplitude of the two-excited state rapidly increases when the nonreciprocal coupling is smaller than the optimal value. On the other hand, we find that the fluctuation of nonreciprocal coupling monotonously changes the CPB locations [see the dashed white and magenta lines in Fig.~\ref{fig:5}(d)].

\begin{figure}
	\includegraphics[width=0.8\linewidth]{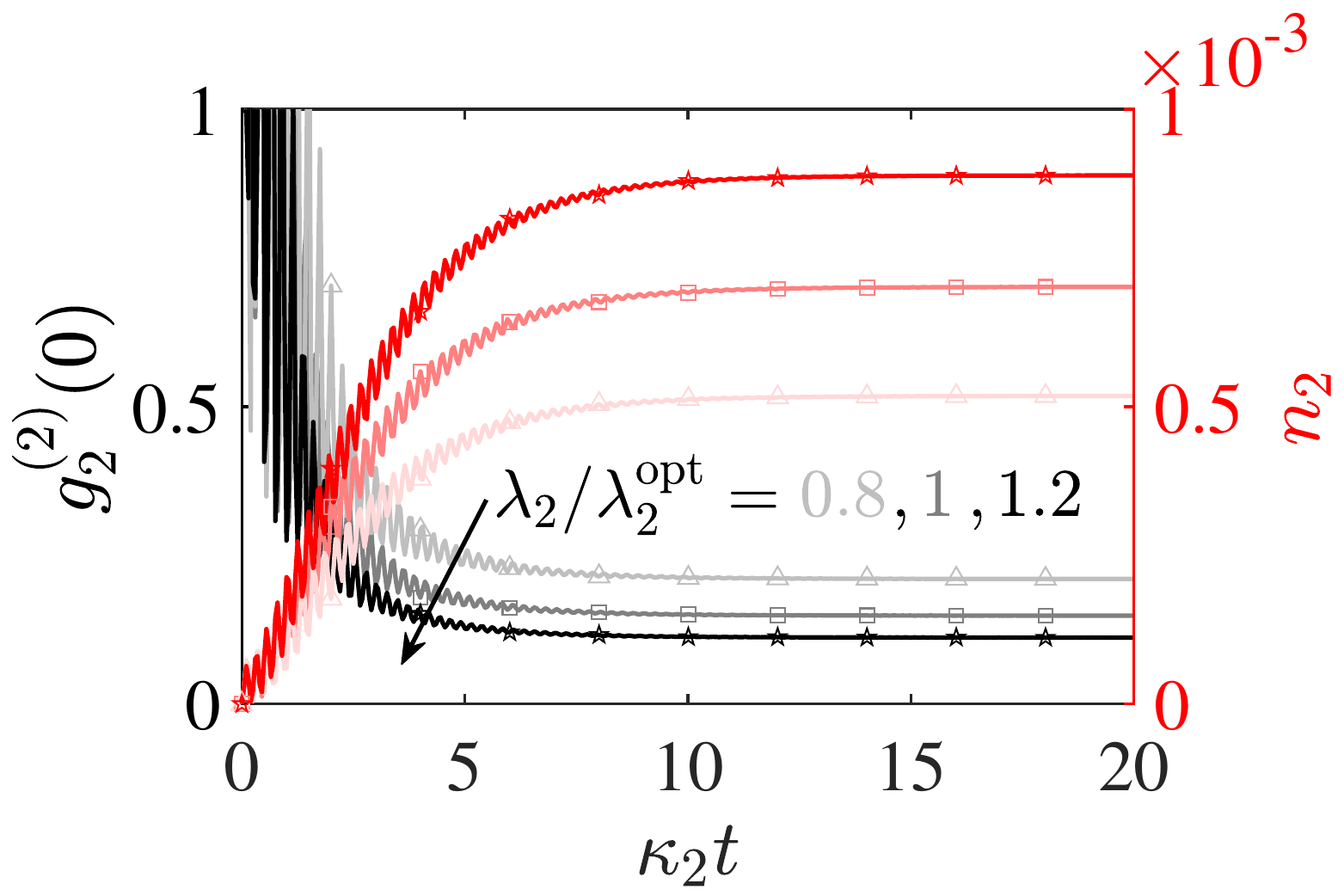}
	\caption{\label{fig:6}The dynamical evolution of the equal-time second-order correlation function and the intracavity photon number with different nonreciprocal couplings. Here, we choose the cavity-laser detuning $\Delta_{2}=\sqrt{\lambda_{1}\lambda_{2}+\chi^{2}/4}+\chi/2$ [the rightmost dip in Fig.~\ref{fig:5}(a)].}
\end{figure}

Moreover, we study the dynamical evolution of the equal-time second-order correlation function and the intracavity photon number, located at $\Delta_{2}=\sqrt{\lambda_{1}\lambda_{2}+\chi^{2}/4}+\chi/2$, with different nonreciprocal couplings to explore their effect on the CPB. The obtained results of both them are shown together in Fig.~\ref{fig:6}, where the different colors respectively represent the correlation function and intracavity photon number, and the gray scale distinguishes those results of different nonreciprocal couplings. Different from the UPB effect, the CPB and intracavity photon number get better with the nonreciprocal coupling increasing; namely, the smaller correlation function and the larger intracavity photon number are obtained when the nonreciprocal coupling is enhanced. That means the efficiency of single-photon emission and the quality of the single-photon source are both improved with the nonreciprocal coupling increasing, which is completely different from the UPB.

\begin{figure}
	\includegraphics[width=0.49\linewidth]{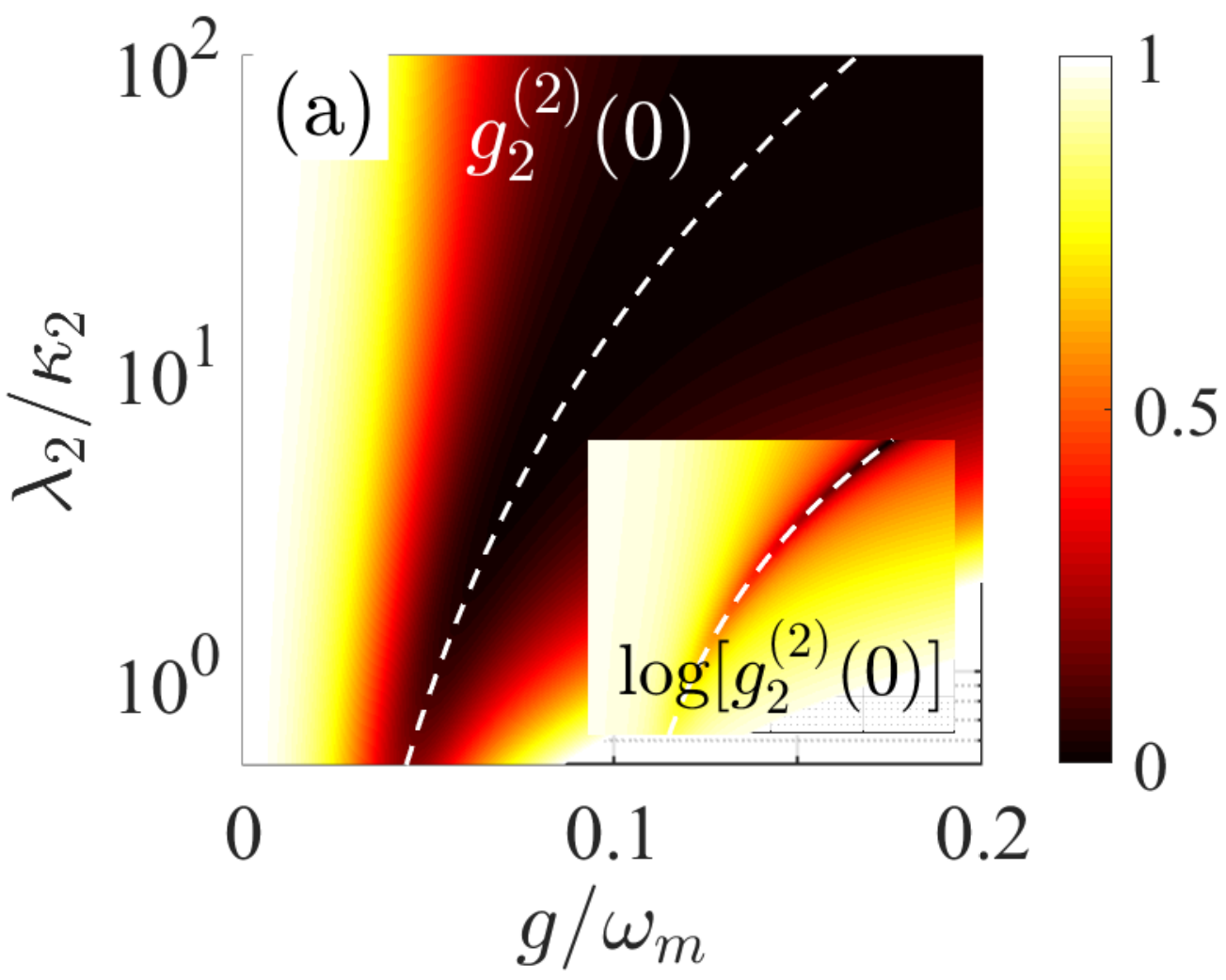}
	\hspace{0in}%
	\includegraphics[width=0.49\linewidth]{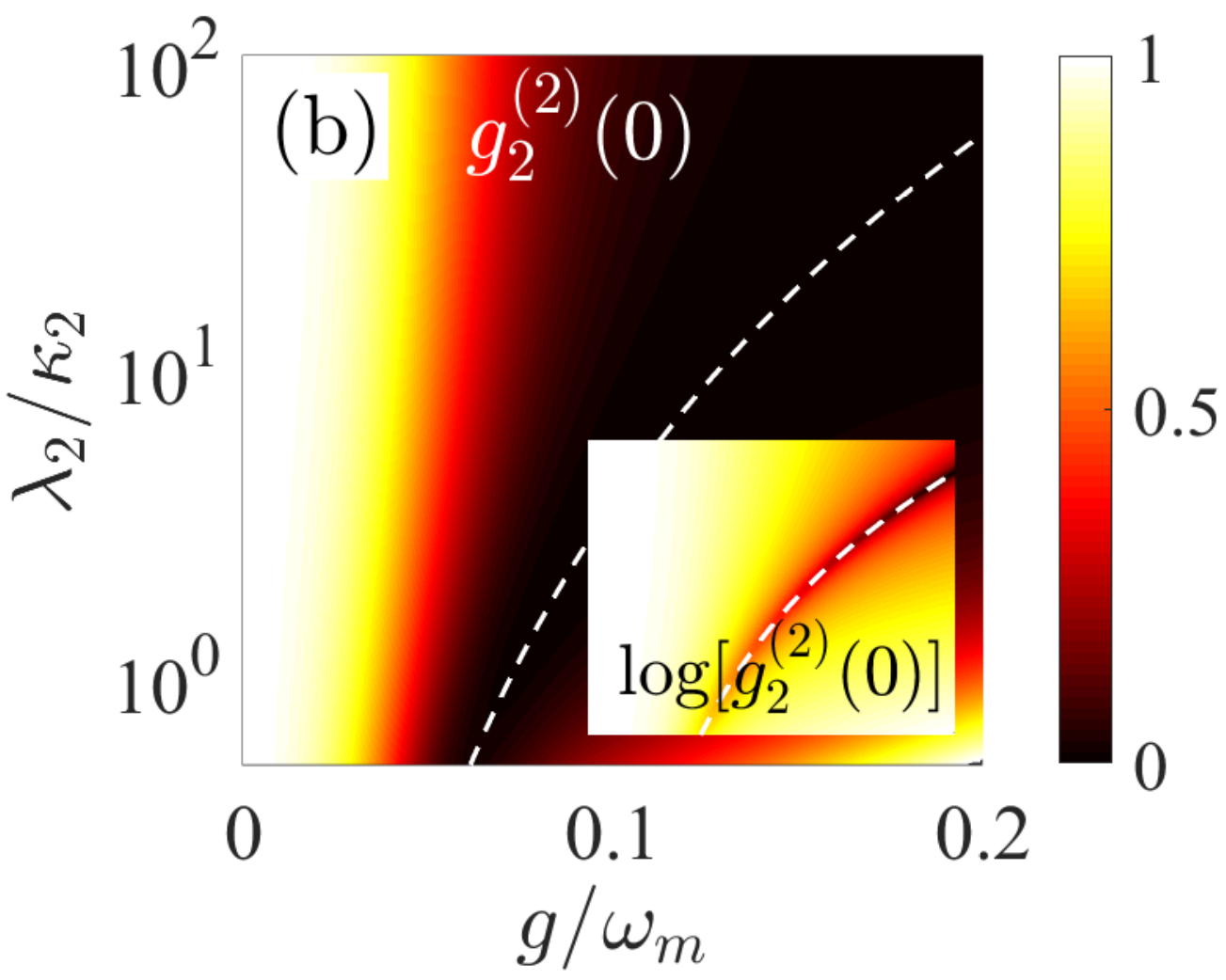}
	\vspace{0in}%
	\includegraphics[width=0.49\linewidth]{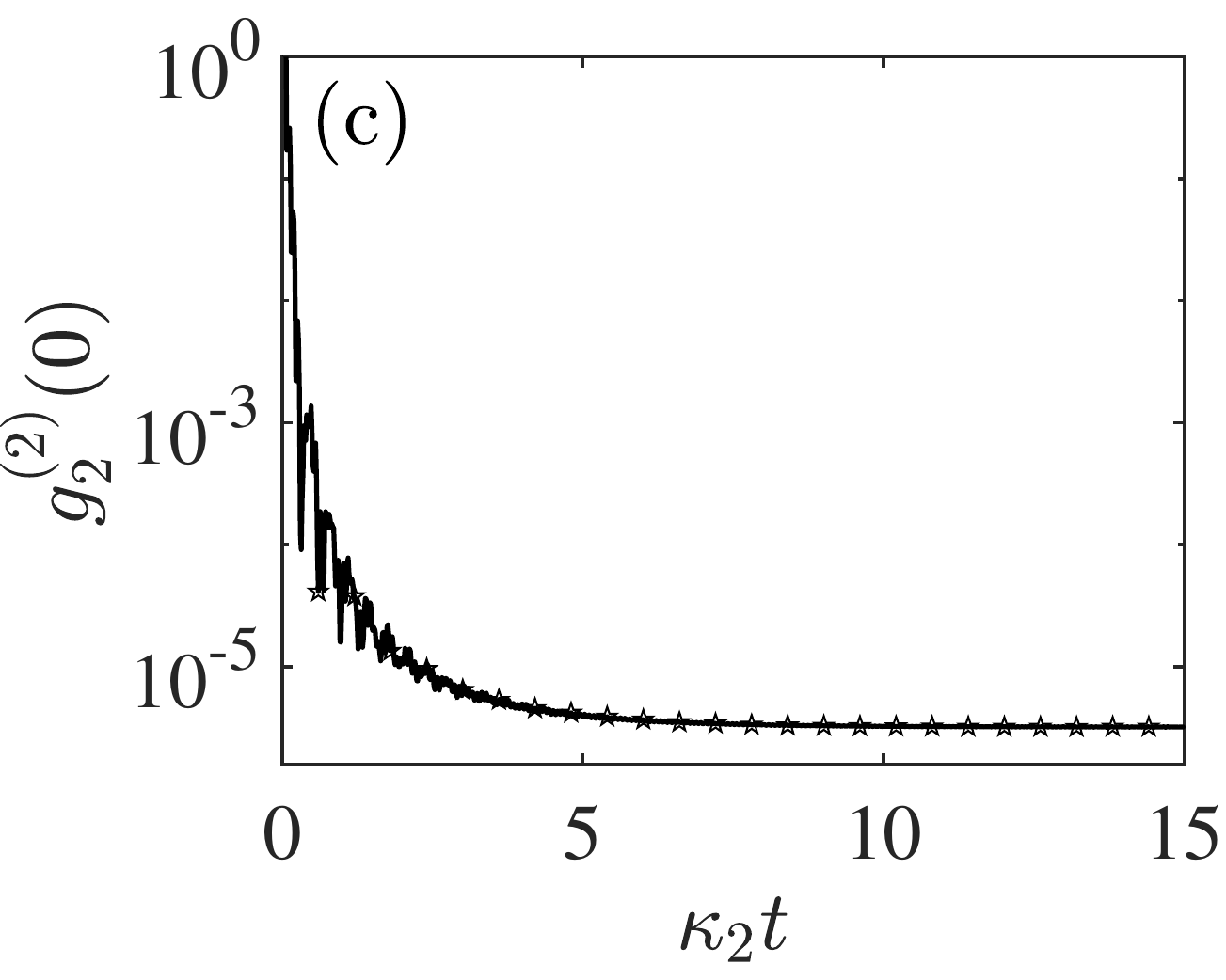}
	\hspace{0in}%
	\includegraphics[width=0.49\linewidth]{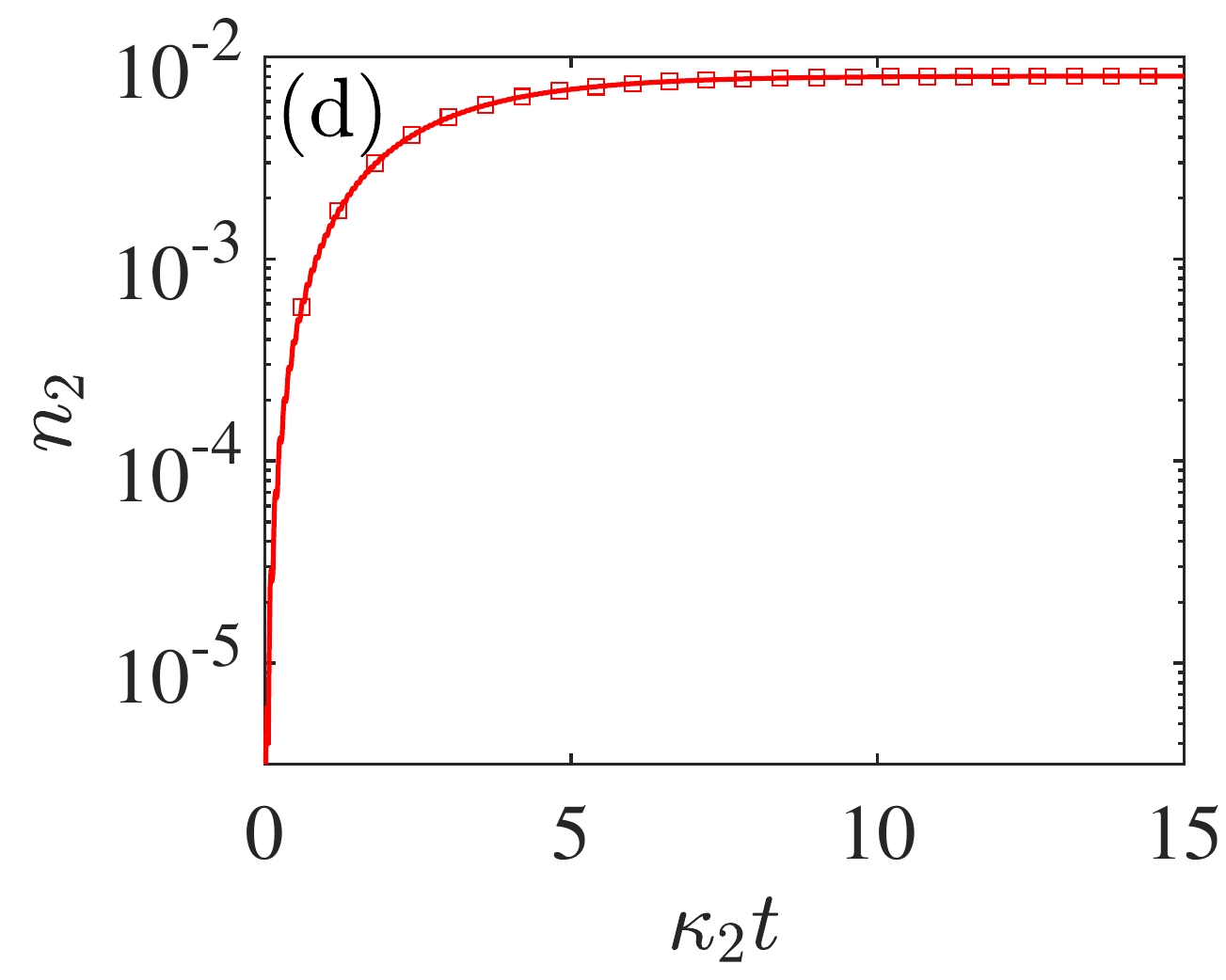}
	\caption{\label{fig:7}(a) and (b) represent the equal-time second-order correlation function $g_{2}^{(2)}(0)$ located at $\Delta_{2}=\sqrt{\lambda_{1}\lambda_{2}+\chi^{2}/4}+\chi/2$ versus both the optomechanical and nonreciprocal couplings when $\lambda_{1}/\kappa_{2}=8$ and $\lambda_{1}/\kappa_{2}=30$, respectively, where the dashed white line is the conventional optimal relation $\lambda_{2}=4.2g^{4}/(\lambda_{1}\omega_{m}^{2})$ and the inserts show the logarithmic value of correlation function $\log[g_{2}^{(2)}(0)]$. (c) and (d) are the dynamical evolutions of the equal-time second-order correlation function and the intracavity photon number located at $\Delta_{2}=\sqrt{\lambda_{1}\lambda_{2}+\chi^{2}/4}+\chi/2$ with $\lambda_{2}/\kappa_{2}=8$.}
\end{figure}

In the above discussions, we still take the unconventional optimal relation [see Eq.~(\ref{e09})] as the standard to study the different blockade behaviors in the strong coupling region. However, it is worth noting that the unconventional optimal relation is not satisfied in the CPB effect, as shown in Fig.~\ref{fig:6}. Naturally, we consider whether there is another optimal relation to make the conventional blockade best. We study those changes of the correlation function versus the optomechanical and nonreciprocal couplings. We then obtain a conventional optimal relation by numerically fitting, $\lambda_{2}=4.2g^{4}/(\lambda_{1}\omega_{m}^2)$, which can optimize the CPB located at $\Delta_{2}=\sqrt{\lambda_{1}\lambda_{2}+\chi^{2}/4}+\chi/2$, as shown in Figs.~\ref{fig:7}(a) and \ref{fig:7}(b). However, it is worth noting that the UPB disappears under the conventional optimal relation due to the destructive quantum interference cannot be satisfied. Moreover, the dynamical evolutions of the equal-time second-order correlation function and the intracavity photon number are also calculated and shown in Figs.~\ref{fig:7}(c) and \ref{fig:7}(d), respectively. We can see that the correlation function is almost zero over time and the intracavity photon number up to $\sim10^{-2}$ in the strong coupling region.

\begin{figure}
	\includegraphics[width=0.8\linewidth]{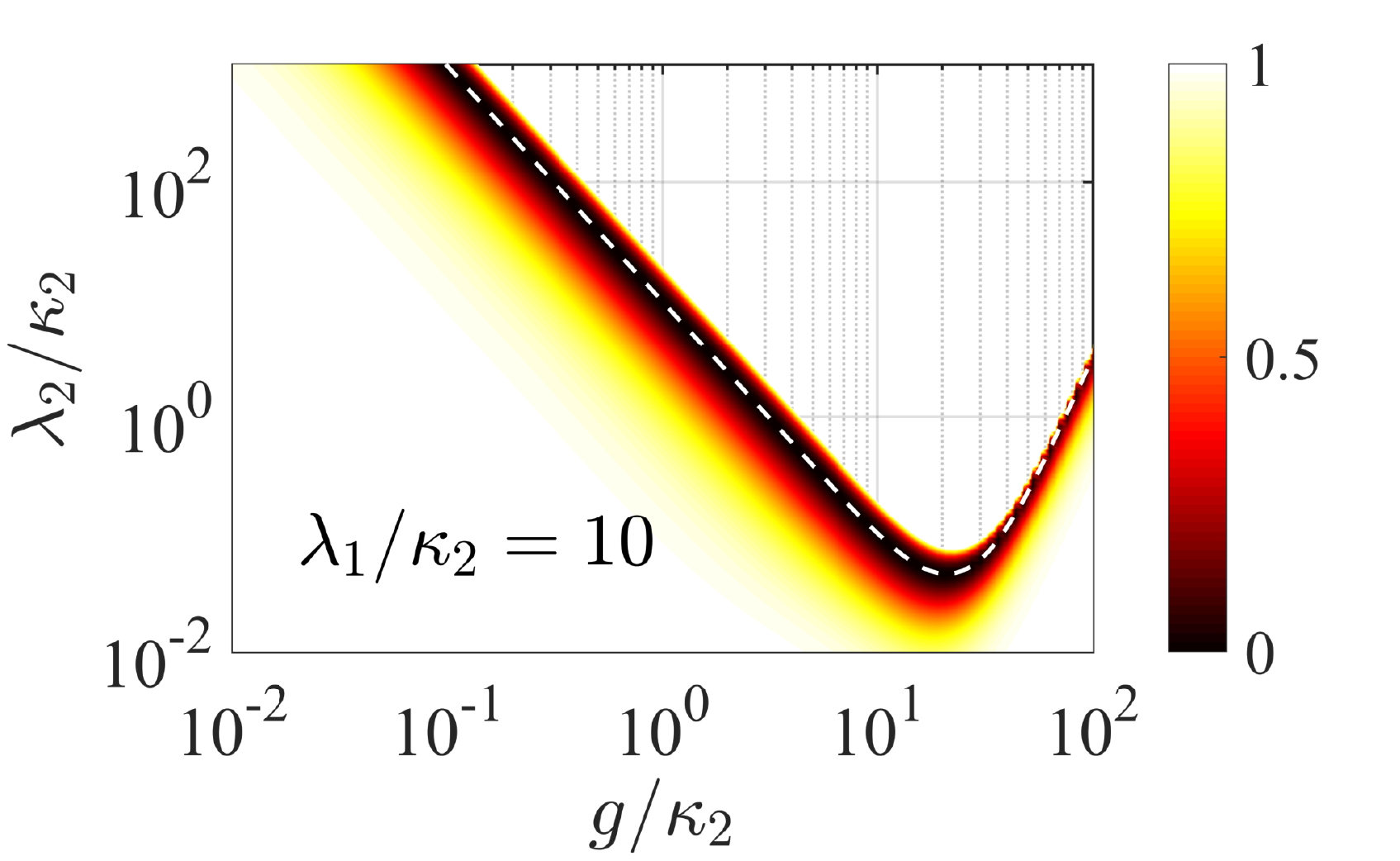}
	\caption{\label{fig:8}The equal-time second-order correlation function versus both the optomechanical coupling and the nonreciprocal coupling. Here, the dashed white line represents the optimal parameter relation in Eq.~(\ref{e09}).}
\end{figure}

On the other hand, the single-photon optomechanical coupling is usually smaller than the optical decay rate in most optomechanical experiments. Therefore, the selected ideal optomechanical coupling $g/\kappa_{2}\simeq21$ is still a challenge in the current experiment. Moreover, the optomechanical coupling is not as large as possible in our work, which is different from the traditional optomechanical system~\cite{PhysRevLett.107.063601,PhysRevLett.107.063602}. Here, we continue to explore the possibility of PB occurring when the optomechanical coupling is smaller than optical damping rate $g<\kappa_{2}$ and show the result in Fig.~\ref{fig:8}. Obviously, when the optomechanical coupling is smaller than the optical damping rate, the strong PB is achieved only in the strong nonreciprocal coupling region, which is coincident with the previous reports~\cite{JPB.46.035502,arXiv:1302.5937}. Although the nonreciprocal coupled auxiliary cavity reduces the requirement of optomechanical coupling, the ideal optomechanical coupling strength is still the goal pursued by researchers in the future.

\section{\label{sec.4}Conclusions}
In conclusion, we have investigated the photon anti-bunching effect in a double-cavity optomechanical system with nonreciprocal coupling via calculating the second-order correlation function analytically and numerically. Different blockade mechanisms are distinguished and discussed both in the weak and strong coupling regions. Specifically, the CPB only occurs in the strong coupling region, while the UPB can survive even in the weak couping region ($\{\lambda_{1},\lambda_{2}\}<\kappa_{2}$ and $g/\omega_{m}\ll1$). To achieve the perfect PB, we respectively derive the conventional and unconventional optimal parameter relations by analyzing the respective blockade mechanisms. Moreover, we give an optomechanical coupling threshold to generate the perfect PB effect in the weak coupling region, and both the smaller and larger optomechanical couplings require a larger nonreciprocal coupling to achieve the same blockade effect. Meanwhile, we find that those PBs under respective mechanisms (CPB and UPB) show completely different behaviors with the change of the nonreciprocal coupling. Furthermore, the dynamical evolutions of second-order correlation function and intracavity photon number are respectively calculated via simulating the quantum master equation numerically, which can further verify all the analyses under the steady-state assumption. Our work paves a way for the experimental implementation of the single-photon source and might be meaningful in generating the few-photon quantum states.

\begin{acknowledgments}
This work was supported by the National Natural Science Foundation of China under Grant No. 61822114, No. 61575055, and No. 11874132.
\end{acknowledgments}

\appendix
\section{\label{app1}Derivation of system Hamiltonian}
The original Hamiltonian of the whole system is given by
\begin{eqnarray}\label{eA01}
H_{1}&=&\omega_{1}a_{1}^{\dagger}a_{1}+\omega_{2}a_{2}^{\dagger}a_{2}+\omega_{m}b^{\dagger}b+\lambda_{1}a_{1}^{\dagger}a_{2}+\lambda_{2}a_{1}a_{2}^{\dagger}\cr\cr
&&-ga_{1}^{\dagger}a_{1}(b^{\dagger}+b)+Ea_{2}^{\dagger}e^{-i\omega_{l}t}+E^{\ast}a_{2}e^{i\omega_{l}t},
\end{eqnarray}
where $\omega_{j}$ is the resonance frequency of the $j$th cavity. By performing a rotating transformation defined by $V=\exp[-i\omega_{l}t(a_{1}^{\dagger}a_{1}+a_{2}^{\dagger}a_{2})]$, the system Hamiltonian is transformed as
\begin{eqnarray}\label{eA02}
H_{2}&=&\Delta_{1}a_{1}^{\dagger}a_{1}+\Delta_{2}a_{2}^{\dagger}a_{2}+\omega_{m}b^{\dagger}b+\lambda_{1}a_{1}^{\dagger}a_{2}+\lambda_{2}a_{1}a_{2}^{\dagger}\cr\cr
&&-ga_{1}^{\dagger}a_{1}(b^{\dagger}+b)+Ea_{2}^{\dagger}+E^{\ast}a_{2},
\end{eqnarray}
which is Eq.~(\ref{e01}) in the main text. Transforming into the mechanical displacement representation by a canonical transformation $V^{\prime}=\exp[g/\omega_{m}a_{1}^{\dagger}a_{1}(b^{\dagger}-b)]$, the Hamiltonian changes
\begin{eqnarray}\label{eA03}
H_{3}&=&\Delta_{1}a_{1}^{\dagger}a_{1}+\Delta_{2}a_{2}^{\dagger}a_{2}+\omega_{m}b^{\dagger}b\cr\cr
&&+\lambda_{1}a_{1}^{\dagger}a_{2}e^{-\frac{g}{\omega_{m}}a_{1}^{\dagger}a_{1}(b^{\dagger}-b)}+\lambda_{2}a_{1}a_{2}^{\dagger}e^{\frac{g}{\omega_{m}}a_{1}^{\dagger}a_{1}(b^{\dagger}-b)}\cr\cr
&&-\frac{g^{2}}{\omega_{m}}(a_{1}^{\dagger}a_{1})^{2}+Ea_{2}^{\dagger}+E^{\ast}a_{2}.
\end{eqnarray}
Due to the weak optomechanical coupling ($g/\omega_{m}\ll1$), the exponential factors in Eq.~(\ref{eA03}) can be approximately omitted. The system Hamiltonian then is rewritten as
\begin{eqnarray}\label{eA04}
H_{4}&=&\Delta_{1}a_{1}^{\dagger}a_{1}+\Delta_{2}a_{2}^{\dagger}a_{2}+\omega_{m}b^{\dagger}b+\lambda_{1}a_{1}^{\dagger}a_{2}+\lambda_{2}a_{1}a_{2}^{\dagger}\cr\cr
&&-\frac{g^{2}}{\omega_{m}}(a_{1}^{\dagger}a_{1})^{2}+Ea_{2}^{\dagger}+E^{\ast}a_{2}.
\end{eqnarray}
We can see that the mechanical resonator is decoupled with the optical cavity, which means the evolutions of optical and mechanical parts are independent each other, i.e., the state evolution of the total system $e^{-iH_{4}t}|\psi\rangle_{\mathrm{system}}=e^{-iH_{5}t}|\psi\rangle_{\mathrm{optical}}\otimes e^{-i\omega_{m}tb^{\dagger}b}|\psi\rangle_{\mathrm{mechanical}}$. When we study the photon statistic in the system, the mechanical part in Eq.~(\ref{eA04}) can be ignored safely. The Hamiltonian then becomes
\begin{eqnarray}\label{eA05}
H_{5}&=&\Delta_{1}a_{1}^{\dagger}a_{1}+\Delta_{2}a_{2}^{\dagger}a_{2}+\lambda_{1}a_{1}^{\dagger}a_{2}+\lambda_{2}a_{1}a_{2}^{\dagger}\cr\cr
&&-\frac{g^{2}}{\omega_{m}}(a_{1}^{\dagger}a_{1})^{2}+Ea_{2}^{\dagger}+E^{\ast}a_{2},
\end{eqnarray}
which is Eq.~(\ref{e02}) in the main text when $\chi=g^{2}/\omega_{m}$.
\section{\label{app2}Driving the optomechanical cavity}
In the main text,  the classical laser field used in the model system is to drive the second cavity, in which the perfect PB can be observed with the optimal parameters relations in Eq.~(\ref{e09}). Here, as a contrast, we simply discuss the photon statistic when the classical laser field drives the optomechanical cavity. Meanwhile, the reduced Hamiltonian is given by
\begin{eqnarray}\label{eB01}
H_{6}&=&\Delta_{1}a_{1}^{\dagger}a_{1}+\Delta_{2}a_{2}^{\dagger}a_{2}+\lambda_{1}a_{1}^{\dagger}a_{2}+\lambda_{2}a_{1}a_{2}^{\dagger}\cr\cr
&&-\frac{g^{2}}{\omega_{m}}(a_{1}^{\dagger}a_{1})^{2}+Ea_{1}^{\dagger}+E^{\ast}a_{1}.
\end{eqnarray}
Similar to the calculation in the main text, the steady-state probability amplitudes can be obtained
\begin{eqnarray}\label{eB02}
C_{10}&=&2E\left(2\Delta_{2}-i\kappa_{2}\right)/M,\cr\cr
C_{01}&=&-4E\lambda_{2}/M,\cr\cr
C_{20}&=&2\sqrt{2}E^{2}\left(2\Delta_{2}-i\kappa_{2}\right)^{2}\left(\chi-2\Delta_{2}+i\kappa_{2}\right)/N,\cr\cr
C_{11}&=&8E^{2}\lambda_{2}\left(2\Delta_{2}-i\kappa_{2}\right)\left(2\Delta_{2}-2\chi-i\kappa_{2}\right)/N,\cr\cr
C_{02}&=&8\sqrt{2}E^{2}\lambda_{2}^{2}\left(2\chi-2\Delta_{2}+i\kappa_{2}\right)/N,
\end{eqnarray}
with
\begin{eqnarray}\label{eB03}
M&=&\big[4\lambda_{1}\lambda_{2}+\left(2\Delta_{2}-i\kappa_{2}\right)\left(2\chi-2\Delta_{2}+i\kappa_{2}\right)\big],\cr\cr
N&=&\big[4\lambda_{1}\lambda_{2}(2\chi-2\Delta_{2}+i\kappa_{2})+(2\Delta_{2}-i\kappa_{2})\cr\cr
&&\times(\chi-2\Delta_{2}+i\kappa_{2})(4\chi-2\Delta_{2}+i\kappa_{2})\big]M.
\end{eqnarray}
It is easy to find that the perfect PB cannot occur in the second cavity due to there is not real solution for equation $|C_{02}|=0$. For the optomechanical cavity, the condition of the perfect PB occurring ($|C_{20}|=0$) is
\begin{eqnarray}\label{eB04}
12\Delta_{2}^{2}-4\chi\Delta_{2}&=&\kappa^{2},\cr\cr
4\Delta_{2}\left(\chi-4\Delta_{2}\right)^{2}&=&0,
\end{eqnarray}
which also cannot be satisfied at the same time in the real space. So the perfect PB cannot occur when the classical laser field drives the optomechanical cavity in the nonreciprocal coupled double-cavity optomechanical system. Therefore, the equal-time second-order correlation function of the optomechanical cavity is
\begin{eqnarray}\label{eB05}
g_{1}^{(2)}(0)=\frac{\left(\chi-2\Delta_{2}\right)^{2}+\kappa_{2}^{2}}{|N/M^{2}|^{2}},
\end{eqnarray}
which again proves the perfect PB is impossible, i.e., $(\chi-2\Delta_{2})^{2}+\kappa_{2}^{2}\neq0$. However, we also need to emphasize that the occurring of PB is possible because it only needs to satisfy $g_{1}^{(2)}(0)\ll1$. We find that the most strong PB can occur in the vicinity of $\Delta_{2}=\chi/2$. Experimentally, the mechanical motion of optical microresonators is related to many factors, such as size~\cite{Nature.482.63}, material~\cite{PhysRevA.82.031804}, external laser pump, etc. For the considered system in the main text, it could be expected to be implemented via asymmetric structural engineering.


\begin{thebibliography}{999}%
\bibitem{PhysRevLett.105.137401}
W.~M. Witzel, A.~Shabaev, C.~S. Hellberg, V.~L. Jacobs, and A.~L. Efros, Phys.
Rev. Lett. \textbf{105}, 137401 (2010).

\bibitem{RevModPhys.81.1301}
V.~Scarani, H.~Bechmann-Pasquinucci, N.~J. Cerf, M.~Du\ifmmode~\check{s}\else
\v{s}\fi{}ek, N.~L\"utkenhaus, and M.~Peev, Rev. Mod. Phys. \textbf{81},
1301 (2009).

\bibitem{PhysRevA.72.052330}
L.~Childress, J.~M. Taylor, A.~S. S\o{}rensen, and M.~D. Lukin, Phys. Rev. A
\textbf{72}, 052330 (2005).

\bibitem{PhysRevLett.114.170802}
K.~R. Motes, J.~P. Olson, E.~J. Rabeaux, J.~P. Dowling, S.~J. Olson, and P.~P.
Rohde, Phys. Rev. Lett. \textbf{114}, 170802 (2015).

\bibitem{PhysRevLett.79.1467}
A.~Imamo\ifmmode~\bar{g}\else \={g}\fi{}lu, H.~Schmidt, G.~Woods, and
M.~Deutsch, Phys. Rev. Lett. \textbf{79}, 1467 (1997).

\bibitem{PhysRevA.49.R20}
W.~Leo\ifmmode~\acute{n}\else \'{n}\fi{}ski and R.~Tana\ifmmode~\acute{s}\else
\'{s}\fi{}, Phys. Rev. A \textbf{49}, R20 (1994).

\bibitem{PhysRevA.46.R6801}
L.~Tian and H.~J. Carmichael, Phys. Rev. A \textbf{46}, R6801 (1992).

\bibitem{PhysRevLett.104.183601}
T.~C.~H. Liew and V.~Savona, Phys. Rev. Lett. \textbf{104}, 183601 (2010).

\bibitem{PhysRevA.83.021802}
M.~Bamba, A.~Imamo\ifmmode~\breve{g}\else \u{g}\fi{}lu, I.~Carusotto, and
C.~Ciuti, Phys. Rev. A \textbf{83}, 021802(R) (2011).

\bibitem{NJP.15.025014}
M.~Bajcsy, A.~Majumdar, A.~Rundquist, and J.~Vu{\v{c}}kovi{\'{c}}, New J. Phys.
\textbf{15}, 025014 (2013).

\bibitem{Nature.436.87}
K.~M. Birnbaum, A.~Boca, R.~Miller, A.~D. Boozer, T.~E. Northup, and H.~J.
Kimble, Nature (London) \textbf{436}, 87 (2005).

\bibitem{NatPhys.4.859}
A.~Faraon, I.~Fushman, D.~Englund, N.~Stoltz, P.~Petroff, and
J.~Vu\v{c}kovi\'{c}, Nat. Phys. \textbf{4}, 859 (2008).

\bibitem{PhysRevLett.107.053602}
A.~J. Hoffman, S.~J. Srinivasan, S.~Schmidt, L.~Spietz, J.~Aumentado, H.~E.
T\"ureci, and A.~A. Houck, Phys. Rev. Lett. \textbf{107}, 053602 (2011).

\bibitem{PhysRevLett.121.043601}
H.~J. Snijders, J.~A. Frey, J.~Norman, H.~Flayac, V.~Savona, A.~C. Gossard,
J.~E. Bowers, M.~P. van Exter, D.~Bouwmeester, and W.~L\"offler, Phys. Rev.
Lett. \textbf{121}, 043601 (2018).

\bibitem{PhysRevLett.121.043602}
C.~Vaneph, A.~Morvan, G.~Aiello, M.~F\'echant, M.~Aprili, J.~Gabelli, and
J.~Est\`eve, Phys. Rev. Lett. \textbf{121}, 043602 (2018).

\bibitem{PhysRevB.85.033303}
S.~Ferretti and D.~Gerace, Phys. Rev. B \textbf{85}, 033303 (2012).

\bibitem{PhysRevB.87.235319}
A.~Majumdar and D.~Gerace, Phys. Rev. B \textbf{87}, 235319 (2013).

\bibitem{PhysRevA.99.043837}
F.~Zou, L. B. Fan, J. F. Huang, and J. Q. Liao, Phys. Rev. A \textbf{99},
043837 (2019).

\bibitem{JPA.51.095302}
X.~Y. Zhang, Y.~H. Zhou, Y.~Q. Guo, and X.~X. Yi, J. Phys. A: Math. Theor.
\textbf{51}, 095302 (2018).

\bibitem{PhysRevA.82.032101}
Y. X. Liu, A.~Miranowicz, Y.~B. Gao, J.~Bajer, C.~P. Sun, and F.~Nori, Phys.
Rev. A \textbf{82}, 032101 (2010).

\bibitem{PhysRevA.87.023809}
A.~Miranowicz, M.~Paprzycka, Y. X. Liu, J.~Bajer, and F.~Nori, Phys. Rev. A
\textbf{87}, 023809 (2013).

\bibitem{PhysRevA.99.063828}
I.~Pietik\"ainen, J.~Tuorila, D.~S. Golubev, and G.~S. Paraoanu, Phys. Rev. A
\textbf{99}, 063828 (2019).

\bibitem{PhysRevX.10.021022}
D.~Roberts and A.~A. Clerk, Phys. Rev. X \textbf{10}, 021022 (2020).

\bibitem{PhysRevLett.108.183601}
A.~Majumdar, M.~Bajcsy, A.~Rundquist, and J.~Vu\ifmmode \check{c}\else
\v{c}\fi{}kovi\ifmmode~\acute{c}\else \'{c}\fi{}, Phys. Rev. Lett.
\textbf{108}, 183601 (2012).

\bibitem{PhysRevLett.109.193602}
A.~Ridolfo, M.~Leib, S.~Savasta, and M.~J. Hartmann, Phys. Rev. Lett.
\textbf{109}, 193602 (2012).

\bibitem{PhysRevLett.118.133604}
C.~Hamsen, K.~N. Tolazzi, T.~Wilk, and G.~Rempe, Phys. Rev. Lett. \textbf{118},
133604 (2017).

\bibitem{PhysRevLett.122.243602}
R.~Trivedi, M.~Radulaski, K.~A. Fischer, S.~Fan, and J.~Vu\ifmmode
\check{c}\else \v{c}\fi{}kovi\ifmmode~\acute{c}\else \'{c}\fi{}, Phys. Rev.
Lett. \textbf{122}, 243602 (2019).

\bibitem{PhysRevApplied.12.044065}
J.~Tang, Y.~Deng, and C.~Lee, Phys. Rev. Appl. \textbf{12}, 044065 (2019).

\bibitem{PhysRevA.98.023856}
H.~Z. Shen, C.~Shang, Y.~H. Zhou, and X.~X. Yi, Phys. Rev. A \textbf{98},
023856 (2018).

\bibitem{PhysRevA.100.033814}
J.~Li, C.~Ding, and Y.~Wu, Phys. Rev. A \textbf{100}, 033814 (2019).

\bibitem{PhysRevA.100.043831}
Y. P. Gao, X. F. Liu, T. J. Wang, C.~Cao, and C.~Wang, Phys. Rev. A
\textbf{100}, 043831 (2019).

\bibitem{PhysRevB.100.134421}
Z. X. Liu, H.~Xiong, and Y.~Wu, Phys. Rev. B \textbf{100}, 134421 (2019).

\bibitem{PhysRevA.96.053827}
B.~Sarma and A.~K. Sarma, Phys. Rev. A \textbf{96}, 053827 (2017).

\bibitem{OE.25.6767}
W. W. Deng, G. X. Li, and H.~Qin, Opt. Express \textbf{25}, 6767 (2017).

\bibitem{PhysRevA.100.053857}
A.~Kowalewska-Kud\l{}aszyk, S.~I. Abo, G.~Chimczak,
J.~Pe\ifmmode~\check{r}\else \v{r}\fi{}ina, F.~Nori, and A.~Miranowicz, Phys.
Rev. A \textbf{100}, 053857 (2019).

\bibitem{PhysRevA.95.043838}
H.~Flayac and V.~Savona, Phys. Rev. A \textbf{95}, 043838 (2017).

\bibitem{PhysRevA.92.023838}
Y.~H. Zhou, H.~Z. Shen, and X.~X. Yi, Phys. Rev. A \textbf{92}, 023838 (2015).

\bibitem{PhysRevA.91.063808}
H.~Z. Shen, Y.~H. Zhou, and X.~X. Yi, Phys. Rev. A \textbf{91}, 063808 (2015).

\bibitem{OE.23.32835}
H.~Z. Shen, Y.~H. Zhou, H.~D. Liu, G.~C. Wang, and X.~X. Yi, Opt. Express
\textbf{23}, 32835 (2015).

\bibitem{PhysRevLett.107.063601}
P.~Rabl, Phys. Rev. Lett. \textbf{107}, 063601 (2011).

\bibitem{PhysRevLett.107.063602}
A.~Nunnenkamp, K.~B\o{}rkje, and S.~M. Girvin, Phys. Rev. Lett. \textbf{107},
063602 (2011).

\bibitem{SC.62.970311}
C. H. Bai, D. Y. Wang, S.~Zhang, and H. F. Wang, Sci. China-Phys. Mech. Astron.
\textbf{62}, 970311 (2019).

\bibitem{OL.43.2050}
T. S. Yin, X. Y. L\"{u}, L. L. Wan, S. W. Bin, and Y.~Wu, Opt. Lett.
\textbf{43}, 2050 (2018).

\bibitem{PhysRevLett.109.013603}
K.~Stannigel, P.~Komar, S.~J.~M. Habraken, S.~D. Bennett, M.~D. Lukin,
P.~Zoller, and P.~Rabl, Phys. Rev. Lett. \textbf{109}, 013603 (2012).

\bibitem{PhysRevA.87.013839}
P.~K\'om\'ar, S.~D. Bennett, K.~Stannigel, S.~J.~M. Habraken, P.~Rabl,
P.~Zoller, and M.~D. Lukin, Phys. Rev. A \textbf{87}, 013839 (2013).

\bibitem{PhysRevA.98.013826}
B.~Sarma and A.~K. Sarma, Phys. Rev. A \textbf{98}, 013826 (2018).

\bibitem{JPB.46.035502}
X. W. Xu and Y. J. Li, J. Phys. B: At., Mol. Opt. Phys. \textbf{46}, 035502
(2013).

\bibitem{arXiv:1302.5937}
V. Savona, arXiv:1302.5937 (2013).

\bibitem{OL.45.2604}
D. Y. Wang, C. H. Bai, X.~Han, S.~Liu, S.~Zhang, and H. F. Wang, Opt. Lett.
\textbf{45}, 2604 (2020).

\bibitem{PhysRevA.96.053810}
H.~Flayac and V.~Savona, Phys. Rev. A \textbf{96}, 053810 (2017).

\bibitem{RepProgPhys.70.947}
C.~M. Bender, Rep. Prog. Phys. \textbf{70}, 947 (2007).

\bibitem{Science.363.eaar7709}
M. A. Miri and A.~Al{\`u}, Science \textbf{363}, eaar7709 (2019).

\bibitem{PhysRevLett.80.5243}
C.~M. Bender and S.~Boettcher, Phys. Rev. Lett. \textbf{80}, 5243 (1998).

\bibitem{IEEEJQE.46.1626}
L.~{He}, S.~K. {\"Ozdemir}, Y.~{Xiao}, and L.~{Yang}, IEEE J. Quantum Electron.
\textbf{46}, 1626 (2010).

\bibitem{Nat.Phys.10.394}
B.~Peng, {\c S}.~K. \"Ozdemir, F.~Lei, F.~Monifi, M.~Gianfreda, G.~L. Long,
S.~Fan, F.~Nori, C.~M. Bender, and L.~Yang, Nat. Phys. \textbf{10}, 394
(2014).

\bibitem{Science.346.328}
B.~Peng, {\c S}.~K. {\"O}zdemir, S.~Rotter, H.~Yilmaz, M.~Liertzer, F.~Monifi,
C.~M. Bender, F.~Nori, and L.~Yang, Science \textbf{346}, 328 (2014).

\bibitem{PhysRevA.92.053837}
J.~Li, R.~Yu, and Y.~Wu, Phys. Rev. A \textbf{92}, 053837 (2015).

\bibitem{PhysRevA.100.053820}
J.~Pe\ifmmode\check{r}\else\v{r}\fi{}ina~Jr., A.~Luk\ifmmode~\check{s}\else
\v{s}\fi{}, J.~K. Kalaga, W.~Leo\ifmmode~\acute{n}\else \'{n}\fi{}ski, and
A.~Miranowicz, Phys. Rev. A \textbf{100}, 053820 (2019).

\bibitem{PhysRevLett.121.086803}
S.~Yao and Z.~Wang, Phys. Rev. Lett. \textbf{121}, 086803 (2018).

\bibitem{PhysRevB.22.2099}
W.~P. Su, J.~R. Schrieffer, and A.~J. Heeger, Phys. Rev. B \textbf{22},
2099 (1980).

\bibitem{PhysRevB.97.045106}
S.~Lieu, Phys. Rev. B \textbf{97}, 045106 (2018).

\bibitem{PhysRevA.97.052115}
C.~Yin, H.~Jiang, L.~Li, R.~L\"u, and S.~Chen, Phys. Rev. A \textbf{97}, 052115
(2018).

\bibitem{PhysRevA.94.013815}
H.~Flayac and V.~Savona, Phys. Rev. A \textbf{94}, 013815 (2016).

\bibitem{PhysRevA.99.043818}
D. Y. Wang, C. H. Bai, S.~Liu, S.~Zhang, and H. F. Wang, Phys. Rev. A
\textbf{99}, 043818 (2019).

\bibitem{PhysRevLett.121.153601}
R.~Huang, A.~Miranowicz, J. Q. Liao, F.~Nori, and H.~Jing, Phys. Rev. Lett.
\textbf{121}, 153601 (2018).

\bibitem{PhysRevA.100.053832}
K.~Wang, Q.~Wu, Y. F. Yu, and Z. M. Zhang, Phys. Rev. A \textbf{100}, 053832
(2019).

\bibitem{PhysRevA.101.013826}
H.~Z. Shen, Q.~Wang, J.~Wang, and X.~X. Yi, Phys. Rev. A \textbf{101}, 013826
(2020).

\bibitem{PRJ.7.630}
B.~Li, R.~Huang, X.~Xu, A.~Miranowicz, and H.~Jing, Photon. Res. \textbf{7},
630 (2019).

\bibitem{PhysRevApplied.13.044070}
X. W. Xu, Y.~Li, B.~Li, H.~Jing, and A. X. Chen, Phys. Rev. Applied
\textbf{13}, 044070 (2020).

\bibitem{PhysRevLett.75.4710}
Q.~A. Turchette, C.~J. Hood, W.~Lange, H.~Mabuchi, and H.~J. Kimble, Phys. Rev.
Lett. \textbf{75}, 4710 (1995).

\bibitem{PhysRevLett.77.570}
N.~Hatano and D.~R. Nelson, Phys. Rev. Lett. \textbf{77}, 570 (1996).

\bibitem{PhysRevB.92.094204}
S.~Longhi, D.~Gatti, and G.~Della~Valle, Phys. Rev. B \textbf{92}, 094204
(2015).

\bibitem{PhysRevLett.107.173902}
Y.~Shen, M.~Bradford, and J. T. Shen, Phys. Rev. Lett. \textbf{107}, 173902
(2011).

\bibitem{SR.5.13376}
S.~Longhi, D.~Gatti, and G.~Della~Valle, Sci. Rep. \textbf{5}, 13376 (2015).

\bibitem{PhysRevResearch.2.013280}
X.~Zhu, H.~Wang, S.~K. Gupta, H.~Zhang, B.~Xie, M.~Lu, and Y.~Chen, Phys. Rev.
Research \textbf{2}, 013280 (2020).

\bibitem{PhysRevA.88.023853}
J. Q. Liao and F.~Nori, Phys. Rev. A \textbf{88}, 023853 (2013).

\bibitem{PhysRevA.92.033806}
H.~Wang, X.~Gu, Y. X. Liu, A.~Miranowicz, and F.~Nori, Phys. Rev. A
\textbf{92}, 033806 (2015).

\bibitem{PhysRevA.96.013861}
H.~Xie, C. G. Liao, X.~Shang, M. Y. Ye, and X. M. Lin, Phys. Rev. A
\textbf{96}, 013861 (2017).

\bibitem{PhysRevA.99.013804}
L. L. Zheng, T. S. Yin, Q.~Bin, X. Y. L\"u, and Y.~Wu, Phys. Rev. A
\textbf{99}, 013804 (2019).

\bibitem{OL.43.1163}
K.~Cai, Z. W. Pan, R. X. Wang, D.~Ruan, Z. Q. Yin, and G. L. Long, Opt. Lett.
\textbf{43}, 1163 (2018).

\bibitem{OE.27.27649}
C.~Zhai, R.~Huang, H.~Jing, and L. M. Kuang, Opt. Express \textbf{27},
27649 (2019).

\bibitem{PhysRevLett.113.053604}
H.~Jing, S.~K. \"Ozdemir, X. Y. L\"u, J.~Zhang, L.~Yang, and F.~Nori, Phys.
Rev. Lett. \textbf{113}, 053604 (2014).

\bibitem{PhysRevA.80.033821}
Z. Q. Yin, Phys. Rev. A \textbf{80}, 033821 (2009).

\bibitem{PhysRevA.90.053841}
Y.~Guo, K.~Li, W.~Nie, and Y.~Li, Phys. Rev. A \textbf{90}, 053841 (2014).

\bibitem{PhysRevA.91.033818}
Y. C. Liu, Y. F. Xiao, X.~Luan, Q.~Gong, and C.~W. Wong, Phys. Rev. A
\textbf{91}, 033818 (2015).

\bibitem{PhysRevA.92.023856}
Z.~Li, S. L. Ma, and F. L. Li, Phys. Rev. A \textbf{92}, 023856 (2015).

\bibitem{SR.6.38559}
D. Y. Wang, C. H. Bai, H. F. Wang, A. D. Zhu, and S.~Zhang, Sci. Rep.
\textbf{6}, 38559 (2016).

\bibitem{QST.4.024002}
O.~{\v{C}}ernot{\'{\i}}k, C.~Genes, and A.~Dantan, Quantum Sci. 
Technol. \textbf{4}, 024002 (2019).

\bibitem{RPP.83.026401}
J.~Millen, T.~S. Monteiro, R.~Pettit, and A.~N. Vamivakas, Rep. 
Prog. Phys. \textbf{83}, 026401 (2020).

\bibitem{Nature.482.63}
E.~Verhagen, S.~Del\'{e}glise, S.~Weis, A.~Schliesser, and T.~J. Kippenberg,
Nature (London) \textbf{482}, 63 (2012).

\bibitem{PhysRevA.82.031804}
J.~Hofer, A.~Schliesser, and T.~J. Kippenberg, Phys. Rev. A \textbf{82}, 031804
(2010).
\end{thebibliography}
\end{document}